\documentclass{article}

\usepackage{PRIMEarxiv}
\usepackage[utf8]{inputenc} 
\usepackage[T1]{fontenc}  
\usepackage{hyperref}     
\usepackage{url}           
\usepackage{booktabs}      
\usepackage{amsfonts}    
\usepackage{nicefrac}       
\usepackage{microtype}    
\usepackage{lipsum}
\usepackage{fancyhdr}      
\usepackage{graphicx}      
\graphicspath{{media/}}     
\usepackage{amsmath}

\usepackage{amsfonts,amssymb}
\usepackage{graphicx}
\usepackage{subcaption}
\usepackage{setspace}
\usepackage{tocloft}
\usepackage{algorithm}
\usepackage{algpseudocode}
\DeclareMathOperator*{\argmax}{arg\,max}

\usepackage{authblk}

\makeatletter
\renewcommand\@fnsymbol[1]{ }
\makeatother

\newcommand{\correspondingauthor}{%
  \thanks{* Corresponding author: \texttt{bartomeu.poumulet@bsc.es}}%
}
  
\title{Integrating supervised and reinforcement learning for predictive control with an unmodulated pyramid wavefront sensor for adaptive optics}

\author[1,2,3,*]{Bartomeu Pou\correspondingauthor}
\author[4]{Jeffrey Smith}
\author[1]{Eduardo Quinones}
\author[2]{Mario Martin}
\author[3]{Damien Gratadour}

\affil[1]{Barcelona Supercomputing Center, C. Jordi Girona, 29, 08034, Barcelona, Spain}
\affil[2]{Universitat Polit\`ecnica de Catalunya, C. Jordi Girona, 31, 08034, Barcelona, Spain}
\affil[3]{LESIA, Observatoire de Paris, Universit\'e PSL, Sorbonne Universit\'e, Universit\'e Paris Cit\'e, CNRS, 5 place Jules Janssen, 92195 Meudon, France}
\affil[4]{Australian National University, School of Computing, Canberra, Australian Capital Territory, Australia}

\begin{document}
\maketitle

\begin{abstract}
We propose a novel control approach that combines offline supervised learning to address the challenges posed by non-linear phase reconstruction using unmodulated pyramid wavefront sensors (P-WFS) and online reinforcement learning for predictive control. The control approach uses a high-order P-WFS to drive a tip-tilt stage and a high-dimensional mirror concurrently. Simulation results demonstrate that our method outperforms traditional control techniques, showing significant improvements in performance under challenging conditions such as faint stars and poor seeing, and exhibits robustness against variations in atmospheric conditions.
\end{abstract}


\section{Introduction}
One of the key science drivers for future extremely large optical telescopes is the identification of bio-signatures on rocky exoplanets located within the habitable zone of extrasolar planetary systems. This poses a significant challenge due to the high contrast ratio between the brightness of stars and planets \cite{guyon2018extreme}. To overcome this challenge, extreme adaptive optics (ExAO) systems with exquisite phase reconstruction accuracy and operating at high cadence (1-2 kHz) are required. ExAO systems are usually equipped with a tip-tilt (TT) stage, a high-dimensional deformable mirror (HDDM) and a high-sensitivity wavefront sensor (WFS) such as the pyramid wavefront sensor\cite{ragazzoni1996pupil} (P-WFS). However, the P-WFS high sensitivity carries inherent trade-offs, including a reduced dynamic range and challenges related to optical gains \cite{deo2021correlation}, the latter being the decreased sensitivity as the residual phase amplifies. In addition to these technical challenges is the pressing need for predictive control, given the system's intrinsic lag and the continuously changing atmospheric conditions \cite{stahl1995optimization, guyon2017adaptive, males2018ground}.

AO control techniques based on machine learning (ML) with neural networks have garnered significant attention due to their ability to address the challenges of phase reconstruction and predictive control. Most of these methods involve offline supervised learning (SL), where the model learns to map input/output pairs. We explore ML methods for AO control in the subsequent areas:

\begin{itemize}
    \item \textbf{Non-linear reconstruction}. Initial approaches with dense networks \cite{guo2006wavefront, osborn2012using} progressed into more complex architectures that exploit spatial structures, such as convolutional neural networks (CNN) \cite{swanson2018wavefront, suarez2018improving, landman2020nonlinear, wong2023nonlinear, landman2024making} or methods that learn a more refined model of the data's statistics with generative adversarial networks. The training of the latter model with a dataset formed by closed-loop data generated by a simple integrator has demonstrated its effectiveness for an open-loop correction atop a first-stage linear reconstruction integrator working in closed-loop \cite{smith2022enhanced}. However, for a pure closed-loop regime, the output of the model then affects the subsequent input, which causes a distribution mismatch between training and closed-loop data, having a detrimental impact on the performance of the system. For non-linear reconstruction, training with randomised data distributions offers a potential remedy. Importantly, when using a randomised dataset, it has been shown that dividing the loss function by the magnitude of the phase improves the performance of closed-loop operation as most phases observed have a low root mean square (RMS) value \cite{landman2020nonlinear}.
    \item \textbf{Predictive control}. Similarly, early approaches utilised dense networks \cite{jorgenson1994wavefront, lloyd1996spatio, montera1997prediction}, which have since evolved to include architectures like recurrent neural networks that leverage spatial or temporal patterns \cite{swanson2018wavefront, liu2020wavefront, swanson2021closed, chen2021performance}. The distribution mismatch affects prediction tasks as well. In the case of prediction, possible solutions are to include an extra term in the loss to lead the model's output to have statistics seen in the training set \cite{swanson2021closed} or to train online. To the best of our knowledge, there has not been any work showing the feasibility of online training of neural networks with SL for predictive control. However, there has been some work on online training for traditional linear methods \cite{guyon2017adaptive, haffert2021data}. Recent analysis leveraging telemetry data shows that it is still unclear whether neural networks outperform data-driven linear predictors on-sky. \cite{hafeez2022forecasting, wong2021predictive}

    A promising approach to address the mismatch between training and operational data is by using reinforcement learning (RL). RL learns the internals of a controller by trial and error based on a reward function provided with operational data, avoiding the aforementioned issue. Some initial approaches were investigated using WFS-less methods \cite{hu2018build, ke2019self}. With a WFS, RL in AO can be broadly classified into model-based and model-free methodologies. Model-based RL methods learn a data-driven model of the dynamics of the atmosphere, which is then used to either sample actions that maximize the reward \cite{nousiainen2021adaptive} or learn another controller with data sampled from the model \cite{nousiainen2022towards, nousiainen2022advances, nousiainen2024laboratory}. In contrast, Model-free RL avoids explicit system modelling and adjusts its internal parameters directly based on the reward function feedback. Different algorithms have been proposed for model-free RL for AO \cite{landman2020self, landman2021self, pou2022model}. This study focuses on expanding these model-free RL techniques, with a comparative analysis between model-free and model-based RL planned for future research.
\end{itemize}

We present a methodology that combines different ML strategies to address the non-linear nature of ExAO control with a P-WFS. First, we deploy a neural network architecture trained via SL for high-fidelity wavefront reconstruction. Second, we train a more compact neural network online via RL, which uses the information from this reconstruction. This paper is an extension of our previous work \cite{pou2022adaptive}, introducing the following novel contributions:

\begin{itemize}
    \item Effective application of a combined SL and RL approach for non-linear reconstruction and predictive control. The code is available at \cite{code_for_paper}.
    \item A novel demonstration of an RL model capable of concurrently controlling two correction stages (HDDM and TT).
    \item Achieving enhanced performance across both reconstruction and prediction, surpassing the classical linear integrator performance.
    \item Demonstrating superior system performance even in challenging atmospheric scenarios and in the presence of wavefront sensing noise.
    \item Showcasing system robustness amid changing atmospheric conditions.
    \item Validating the viability of real-time implementation.
\end{itemize}

The paper is structured as follows: Section \ref{Background} introduces the P-WFS core concepts together with wavefront reconstruction using a linear approach and sets the foundation for our models using SL and RL. Section \ref{Methods} explains the diverse models incorporated within our ML system. Section \ref{experimental_setting} discusses the experimental setting. In Section \ref{Results}, we present the outcomes of the experiments for each model. Section \ref{conclusions_future_work} draws conclusions from our experiments and outlines directions for future research.

\section{Background}
\label{Background}

The standard way to reconstruct the incoming phase with a P-WFS is to assume a linear relationship between the phase and the pyramid images. Equation \ref{eq:1_interaction} shows a linear relationship established with the so-called interaction matrix, \textit{D}, between the pyramid pixels, \textit{m}, and the phase projected onto the combination of HDDM and TT, \textit{c}, referred to as reconstruction. The interaction matrix is calibrated by pushing and pulling each actuator in the correction stages by a minimal value and seeing its effect on the P-WFS.

\begin{equation}\label{eq:1_interaction}
     m = D c.
\end{equation}

Taking the generalised inverse of the interaction matrix yields the command matrix, $D^{\dagger}$, which can be used to obtain the reconstruction, \textit{c}, as seen in Equation \ref{eq:2_reconstruction}.

\begin{equation}\label{eq:2_reconstruction}
    c = D^{\dagger} m.
\end{equation}

Building upon the fundamental zonal basis approach, one can transition to a global modal basis. For example, one could use Karhuenen-Loève modes to integrate atmospheric statistics into the controller. In this study, we use the $B_{tt}$ basis \cite{ferreira2018numerical}. This basis is based on the geometrical properties of the AO system and is made for systems with two correction stages: one HDDM and one TT. The Btt basis uses the principal components of the HDDM’s influence functions covariance matrix from which the tip-tilt and piston components have been removed. Then, the TT stage modes are appended at the end. This method ensures the HDDM contribution has no TT component and that the piston is filtered out. Finally, the reconstruction is integrated over time. For timestep \textit{t}, the commands are given by Equation \ref{eq:3_integration}.

\begin{equation}\label{eq:3_integration}
    C_t = C_{t-1} + g c_t,
\end{equation}

where $0 < g \leq 1$ is a gain parameter. Nevertheless, this methodology faces two notable challenges:

\begin{enumerate}
    \item The P-WFS exhibits a pronounced non-linear behaviour, potentially rendering this strategy ineffective. To combat this, pyramid modulation is introduced to linearise the signal. While this modulation broadens the linear range, it concurrently attenuates sensitivity. One of such non-linearities is the so-called optical gains \cite{deo2021correlation}, a reduction of sensitivity associated with each mode and varies during operations depending on the phase content.
    \item Temporal error emerges as a delay is introduced in the loop, particularly as a function of loop cadence and gain, \textit{g}.
\end{enumerate}

This paper proposes an ML strategy that collectively addresses the challenges exposed above.

\begin{figure}[h]
    \centering
    \includegraphics[width=0.9\textwidth]{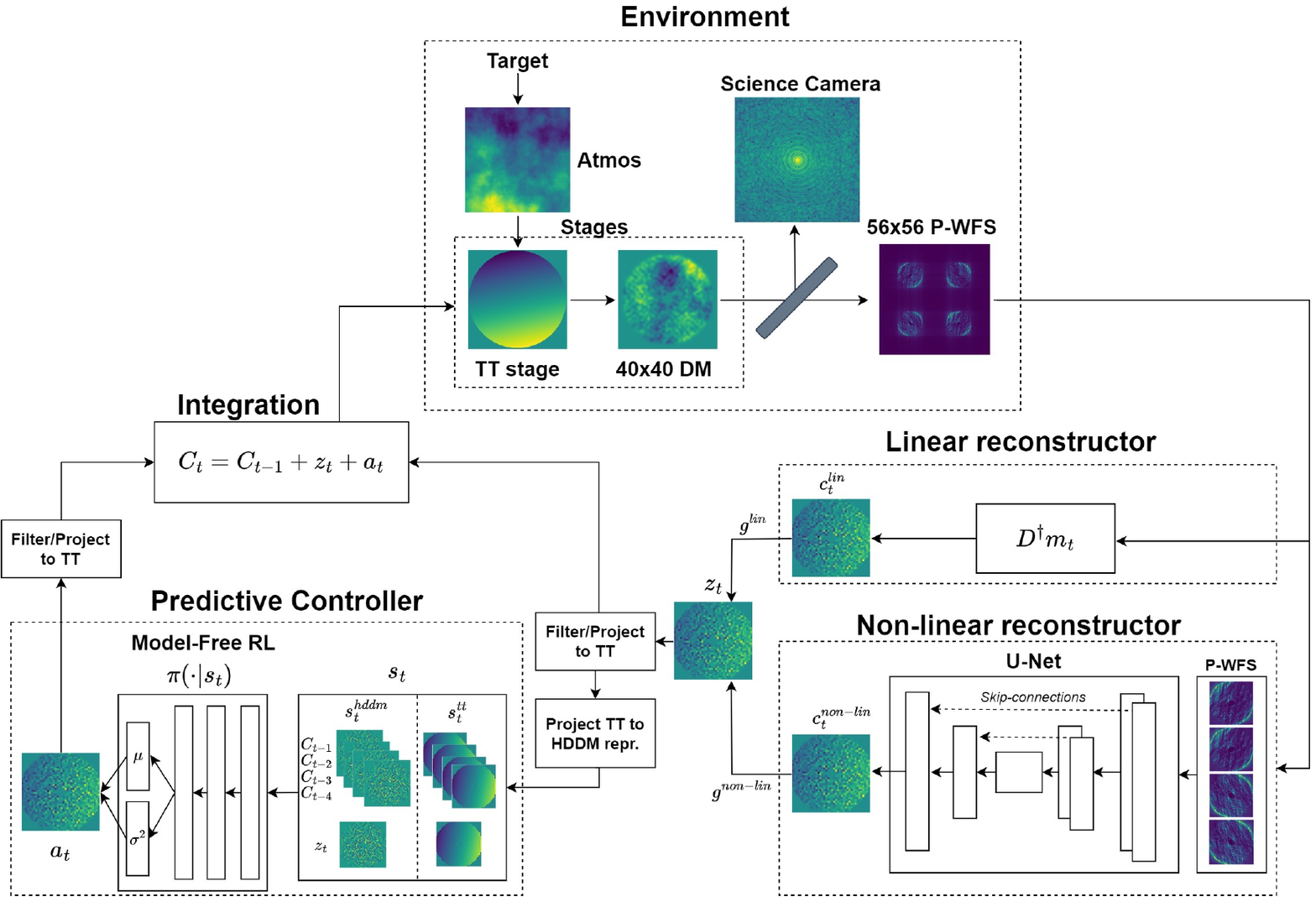}
    \caption{Illustrative diagram of the supervised-reinforcement learning solution: the U-Net is responsible for the non-linear reconstruction, while the RL model takes charge of predictive control. The step of filter/project to TT is explained in Figure \ref{fig:action_rl_unet} and the step of project TT to HDDM representation is explained in Figure \ref{fig:state_reward}.}
    \label{fig:diagram}
\end{figure}

\section{Methods}
\label{Methods}

In this ML solution, the first component employs offline SL to train a U-Net architecture \cite{ronneberger2015u}, specifically for non-linear reconstruction. The subsequent component utilises online RL to train a compact neural network designed for predictive control. This systematic approach is illustrated in Figure \ref{fig:diagram}.

\subsection{First component: non-linear reconstruction with U-Net}

Regarding non-linear phase reconstruction, we use the U-Net due to its demonstrated \textit{state-of-the-art} capabilities for image processing tasks. The U-Net architecture uses CNN layers, which allow for spatial feature extraction. The CNN layers are arranged in a symmetric structure: first, the image is compressed (encoded) into a smaller representation and then decompressed (decoded) back into the original representation. Each layer in the encoding phase has a symmetric layer in the decoding phase connected via skip connections \cite{he2016deep}, a shortcut between layers which bypasses intermediate ones. These connections are crucial for training deep neural networks with a large number of hidden layers. 

\begin{figure}[h]
    \centering
    \includegraphics[width=0.6\textwidth, trim = 0cm 4cm 0cm 3cm, clip]{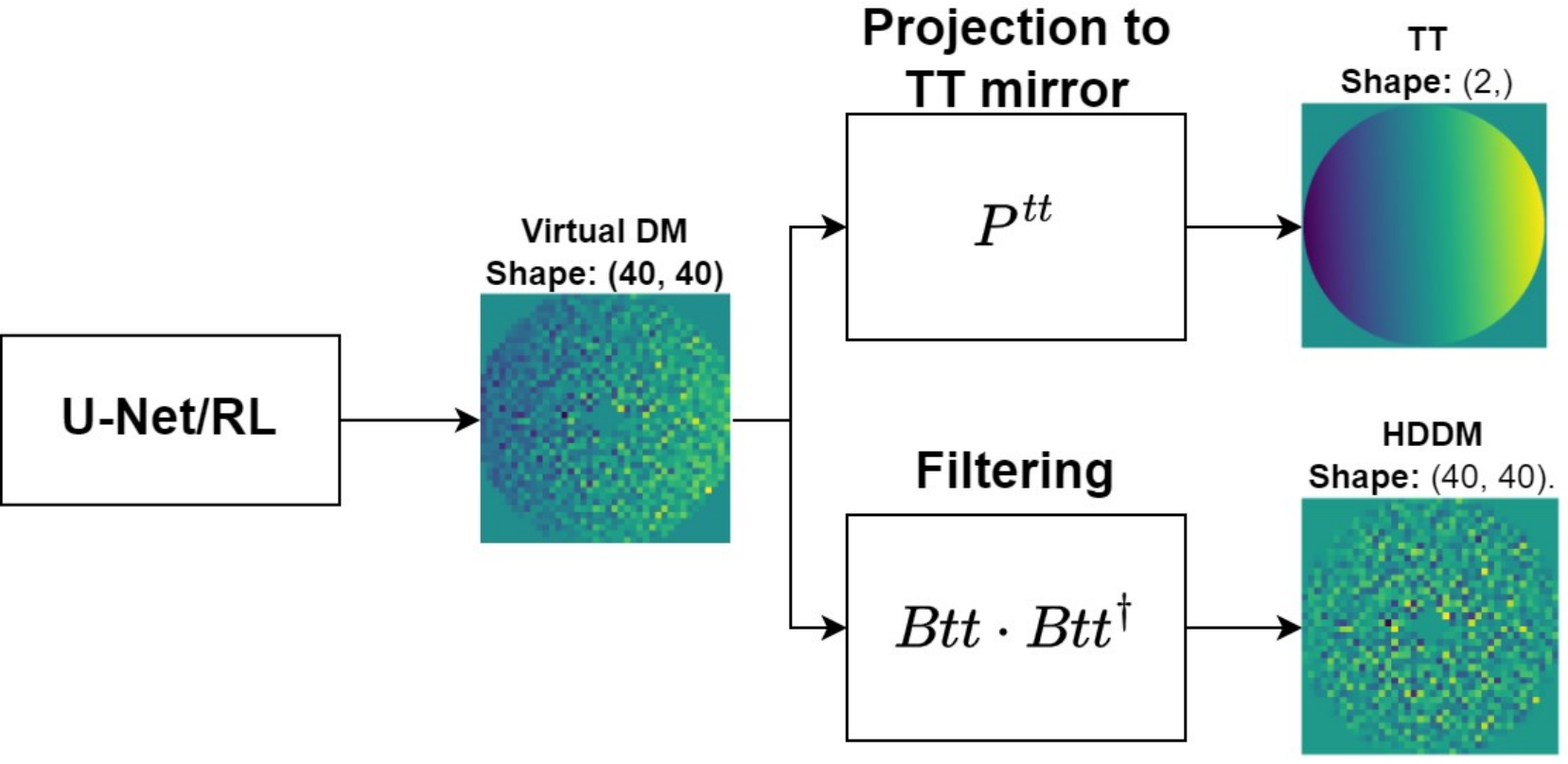}
    \caption{Method to separate the modes for the HDDM and TT, providing the ability to control both stages of correction. $P^{tt}$ is a projection matrix from the virtual DM to TT. $B_{tt}$ is the projection matrix from  $B_{tt}$ basis to actuators and TT stage. $B_{tt}^{\dagger}$ is the generalised inverse of $B_{tt}$.}
    \label{fig:action_rl_unet}
\end{figure}

The input of the U-Net is the photon count from each pixel as measured by the P-WFS. Concretely, we take the P-WFS image and rearrange it into a matrix of size (4, Npix, Npix) where each pyramid face corresponds to a channel and Npix is the number of pixels of a P-WFS face. With that input, the U-Net infers an output matrix which represents the phase projected onto a virtual DM with the same shape as the HDDM containing both HDDM and TT components. To separate them, we do a projection from the virtual DM to TT. The tip-tilt component is then removed by filtering with the $B_{tt}$ modal basis, obtaining the HDDM component. The process of separating TT and HDDM is depicted in Figure \ref{fig:action_rl_unet}.

For efficient closed-loop operations, the non-linear reconstructor must exhibit robust generalisation capabilities. It should predict a phase that emerges from the combination of various processes: the atmospheric disturbance, the commands resulting from prior integrated non-linear reconstructions, and potentially from an RL controller's output. If the training dataset is skewed towards a specific distribution, this can drift off quickly as the non-linear reconstructor model may receive input data with statistics never seen in training. For that, we gather a dataset by randomly shaping the correction stages in our setup. To do so, we sample a value from a log-norm distribution, which is then used as the parameter of a Normal distribution $\mathcal{N}(0, \sigma^2)$ whose output is used to move each actuator. The corresponding dataset is split between training and evaluation. The former is used to learn the U-Net parameters and the latter is used to observe its generalisation capabilities on phases never seen during training.

We use the L1 relative loss as shown in Equation \ref{eq_loss_unet} where $c^*$ is the real reconstructed phase, $c^{non-lin}(m; \theta)$ is the predicted non-linear reconstruction inferred from the P-WFS image pixels, $m$, and processed by the U-Net's weights $\theta$, and $\epsilon$ is a small value to avoid dividing by 0. As seen in the proposed loss, the difference between real and predicted phases is divided by the norm of the real phase. The division by the norm of the real phase is added to keep the gradient within a similar range whatever the phase RMS is, and therefore, avoiding learning a model that focuses on providing good results for phases with high RMS and not for low RMS phases. We opt for L1 relative loss over L2 relative loss due to its resilience against outliers.

\begin{equation}\label{eq_loss_unet}
    L(\theta) = \frac{||c^* - c^{non-lin}(m; \theta)||_1}{||c^*||_1 + \epsilon}.
\end{equation}

As pointed out in the results section in Figure \ref{fig:closed_loop_unet}, the non-linear reconstruction does not ultimately outperform the linear reconstruction alone in all tested atmospheric configurations. For that, at frame $t$, we combine both linear and non-linear reconstructions in $z_t$, as shown in Equation \ref{eq:integrator_unet}.

\begin{equation}\label{eq:integrator_unet}
    z_t(m_t; \theta) = g^{lin} \cdot c_{t}^{lin}(m_t) + g^{non-lin} \cdot c_{t}^{non-lin}(m_t; \theta),
\end{equation}

where $0 \leq g^{lin} \leq 1$, and $0 \leq g^{non-lin} \leq 1$ are the gains for linear and non-linear reconstructions, respectively, and $c_{t}^{lin}$ and $c_{t}^{non-lin}$ are the linear and non-linear reconstructions depending on the P-WFS pixels, $m_t$. The linear combination of reconstructions is designed to leverage the extra accuracy of the non-linear reconstruction while maintaining the robustness of the linear approach.

\subsection{Second component: predictive control with reinforcement learning}

In this section, we discuss our model for predictive control. Our strategy consists of correcting the integrator law (Eq. \ref{eq:3_integration}) built with a combination of linear and non-linear reconstructions (Eq. \ref{eq:integrator_unet}) with a delta correction factor ($\Delta z$) that accounts for the temporal error from delay and atmospheric evolution as seen in Equation \ref{eq:residual}.

\begin{equation}\label{eq:residual}
    C_t(m_t; \theta) = C_{t-1}(m_{t-1}; \theta) + z_t(m_t; \theta) + \Delta z.
\end{equation}

To infer $\Delta z$, we use data-driven methods, specifically RL. RL \cite{sutton2018reinforcement} learns parameters that maximize the return, the cumulative sum of rewards, $r$. At timestep $t$, the return can be expressed as in Equation \ref{return}.

\begin{equation}\label{return}
    R_t = \sum^T_{t'=t} \gamma^{t'-t} r_{t'+1},
\end{equation}

where $\gamma \in [0, 1]$ is a hyperparameter weighing the importance of rewards on future timesteps, and $T$ is the number of steps to finish the task. An RL agent interacts with an environment via trial and error. The overall goal of the RL agent is to learn a function called policy, $\pi$, that maps states, \textit{s}, to actions, \textit{a}, which maximize the return. The optimal policy, $\pi^*$, is expressed mathematically in Equation \ref{eq:4_rl_objective}.

\begin{equation}\label{eq:4_rl_objective}
    \pi^* = \argmax_\pi \mathbb{E}_\pi \big[R_0\big] = \argmax_\pi \mathbb{E}_\pi \left[\sum^T_{t'=0} \gamma^{t'} r_{t'+1}\right],
\end{equation}

where the expectation, $\mathbb{E}_\pi$, is over rewards obtained with a given policy, $\pi$. One common strategy to find the optimal policy via trial and error is by injecting noise into the action and reducing it over time. When the noise is being injected, the policy is in exploration mode. When the noise is discarded, the policy is in exploitation mode. A proven method for intelligently creating this noise is with a slightly modified RL framework: maximum entropy RL \cite{haarnoja2017reinforcement, haarnoja2018soft}. In this modified framework, the actions are considered a random variable, and the policy models a probability distribution over the actions. For a given state, $s$, actions are sampled according to $a \sim \pi(\cdot|s)$, where $\pi(\cdot|s)$ represents the conditional probability distribution of taking any action given the state $s$. The maximum entropy framework uses the entropy of the policy $H(\pi(\cdot|s)) = \mathbb{E}_{a \sim \pi} \big[-\log \pi(a|s)\big]$, where the expectation is over the actions drawn from the policy distribution, which quantifies the amount of noise in the action. Now, not only the cumulative sum of rewards is maximized, but also the entropy of the policy as seen in Equation \ref{eq:5_rl_objective_max_ent}.  While learning, if the rewards obtained are not good enough, the agent focuses on increasing the entropy of the policy, which enforces exploration. When, because of exploration, the agent finds a policy that has a high return, then maximization of Eq. \ref{eq:5_rl_objective_max_ent} automatically reduces the entropy of policy (i.e. reduces exploration) and focuses on maximization of rewards.

\begin{equation}\label{eq:5_rl_objective_max_ent}
    \pi^* = \argmax_\pi \mathbb{E}_\pi \big[R^{MaxEnt}_0\big] =  \argmax_\pi \mathbb{E}_\pi \left[ \sum^T_{t'=0} \gamma^{t'}( r_{t'+1} + \alpha H(\pi(\cdot|s_{t'}))) \right],
\end{equation}

where $\alpha$ is a hyperparameter that balances reward optimization and entropy maximization. The next step is to define $s$, $a$ and $r$ in the specific context of a predictive controller for AO. The state at timestep $t$, $s_t$, is composed of two components: an HDDM representation $s^{hddm}_t$ and a TT representation $s^{tt}_t$. For our architecture, we use CNNs to leverage spatial patterns. For that, each component is structured as 5-channel tensor where each channel is a $Nact \times Nact$ grid where $Nact$ is the number of actuators in the HDDM. This captures HDDM information or the projection of the TT onto the HDDM representation. Each component contains linear and non-linear reconstructions from Equation \ref{eq:integrator_unet}, $z_t^i$, with a list of past commands $C_t^i$ as shown as in Equation \ref{eq:state_1_rl}.

\begin{equation}\label{eq:state_1_rl}
    s^{i}_t = (z_t^{i}, C_{t-1}^{i}, C_{t-2}^{i}, C_{t-3}^{i}, C_{t-4}^{i}); \ \ i \in [hddm, tt].
\end{equation}

where each element is concatenated along a channel dimension. Finally, we concatenate both components in the channel dimension as in Equation \ref{eq:state_2_rl}.

\begin{equation}\label{eq:state_2_rl}
    s_t = (s^{hddm}_t, s^{tt}_t).
\end{equation}

Ultimately, our state adopts a matrix of shape (10, Nact, Nact). The history of past commands provides information to the RL agent to predict the evolution of the atmosphere. We know that including a longer history of commands and a history of residuals can enhance performance, as demonstrated in \cite{nousiainen2022towards}, and remains part of our future research directions. The state could also be formulated as 5 channels by combining the TT and HDDM projections with a single component. However, our initial experiments showed that this representation provides slightly better results. In the separate case, the network is learning features from each type of projection and then combining this information in deeper layers of the network.

With respect to the reward, we must first go through a few concepts. In the context of AO, it is possible to know the system delay to a certain extent. Consequently, we have opted to delay the reward assignment by the intrinsic delay of the system after an action is taken, a strategy we name "delayed assignment". In the case of an AO system that experiences significant jitter, we can assign the reward at the same timestep as we execute the action. The rationale is that RL systems, by nature, optimize for cumulative rewards, so the policy adapts even if it takes more update steps. Unlike the state, we choose not to separate the HDDM and TT components for the reward, as our initial experiments showed this merger leads to better results in the case of the reward. Finally, we introduce an additional term, $p$, which penalizes large gradients between neighbouring command values. We have identified that during training, the RL's underlying trial and error methodology may introduce windup on a few actuators, especially towards the pupil's edges. To avoid this, one possible solution can be to filter more modes on the output of the RL agent, but such an approach can have a strong impact on possible performance improvements from using RL. Alternatively, we can train the RL agent to avoid or recover from such windup by introducing an additional term in the reward. This penalization can take various forms, and we found that the one that achieves better results is one that penalizes higher values of actuators with respect to the values of the neighbour actuators. This penalization can be seen as a regularizer that pushes toward smooth mirror surfaces.

Equation \ref{eq:reward} shows the expression for the reward function of actuator in position $(i, j)$ at timestep $t$ for a delayed assignment of \textit{d} frames, $r_{t+d+1}$, given reconstruction values $z_{t+d+1}^{hddm}$ and $z_{t+d+1}^{tt}$ for the HDDM and TT components and with a penalizing factor on the commands $p_t$ times a scaling parameter $\lambda$. Equation \ref{eq:penalizer} shows the expression for the penalization at position $(i,j)$.

\begin{equation}\label{eq:reward}
    r_{t+d+1}(i, j) = -|z_{t+d+1}^{hddm}(i, j) + z_{t+d+1}^{tt}(i, j)|^2 - \lambda p_t(i,j).
\end{equation}

\begin{equation}\label{eq:penalizer}
    p_t(i,j) = \begin{cases}
    \max_{\substack{(i',j') \in neigh}} \left( \left( C_t(i, j) - C_t(i', j') \right)^2 \right) & \text{if } |C_t(i, j)| > |C_t(i_{\text{max}}, j_{\text{max}})|, \\
    0 & \text{otherwise,}
    \end{cases}
\end{equation}

where $neigh$ is the set of valid neighbour actuators of actuator $(i,j)$ and $(i_{max}, j_{max})$ is the neighbour actuator determined by Equation \ref{eq:argmax}.

\begin{equation} \label{eq:argmax}
    (i_{\text{max}}, j_{\text{max}}) = \argmax_{\substack{(i',j') \in neigh}} \left( \left( C_t(i, j) - C_t(i', j') \right)^2 \right).
\end{equation}

\begin{figure}[h]
    \centering
    \includegraphics[width=0.6\textwidth, trim = 0cm 2cm 0cm 2cm, clip]{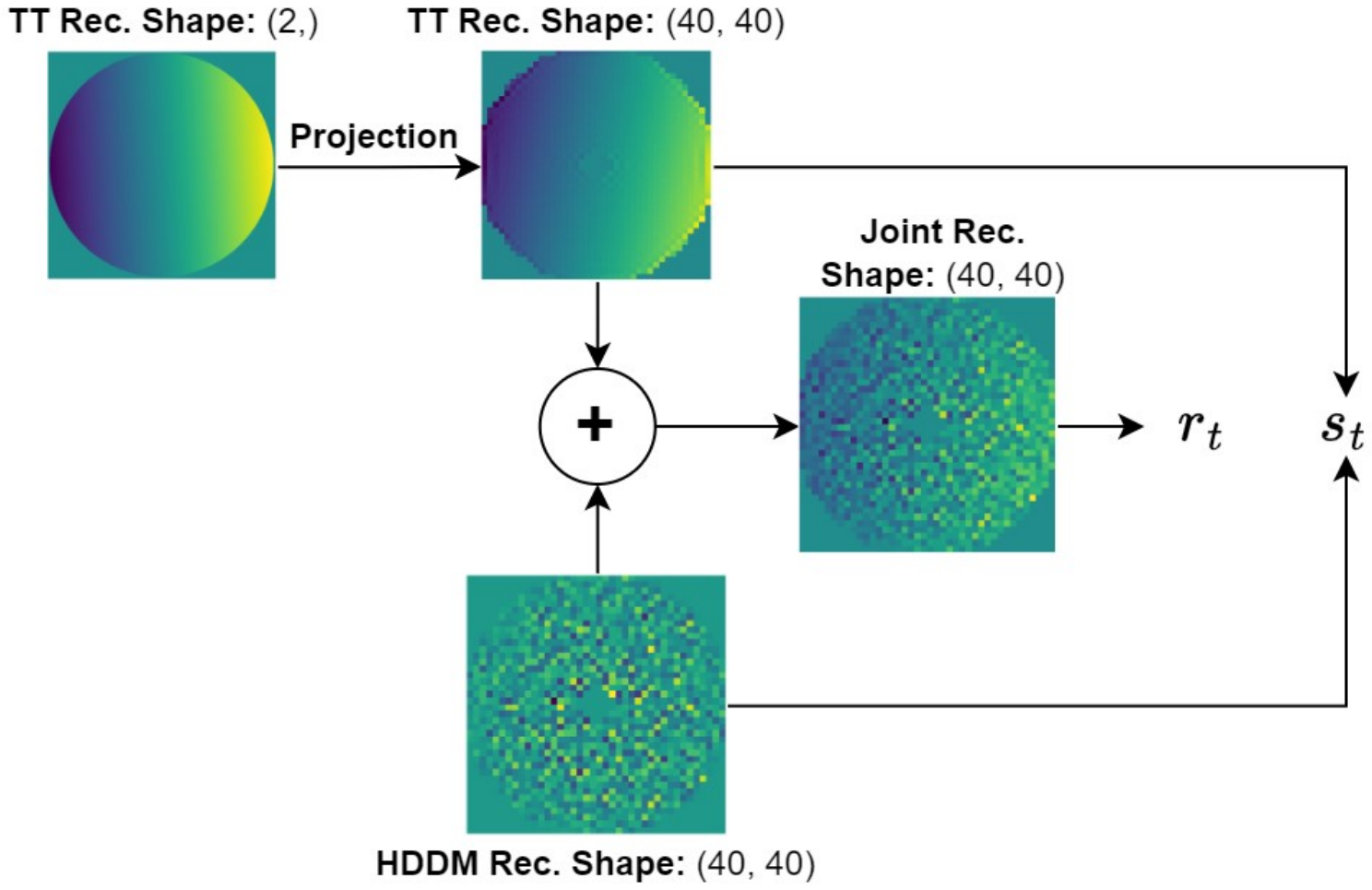}
    \caption{Deriving the state and reward from the reconstruction.}
    \label{fig:state_reward}
\end{figure}

This matrix expression offers a unique sub-reward for each actuator. Given the negative sign in the equation, the design of this reward function prompts policies that strive to minimise the combination of linear and non-linear reconstruction and optimise for a smooth surface in which gradients of neighbour commands do not become too large. For a more comprehensive understanding of the separation of HDDM and TT components of the reconstruction for state and reward, we provide a visual representation as depicted in Figure \ref{fig:state_reward}. The TT component is represented by two values, and the HDDM is represented by the matrix of 40x40 actuators. Once we project the TT into the HDDM representation, we can either sum the components to extract the reward or separate the components and extract the state.

The action is the $\Delta z$ we were looking for. Let the policy be approximated with parameters $\phi$ and the U-Net with parameters $\theta$, the final expression for the command is given in Equation \ref{eq:residual2}.

\begin{equation}\label{eq:residual2}
    C_t(m_t; \theta, \phi_t) = C_{t-1}(m_{t-1}; \theta) + z_t(m_{t}; \theta) + a_t; \ a_t \sim \pi(\cdot|s_t; \phi_t).
\end{equation}

As the weights $\phi$ update online, we must add it a subindex $t$. In RL, the strategy of correcting a basic policy is usually called residual policy learning \cite{silver2018residual, johannink2019residual}. We use Soft Actor-Critic \cite{haarnoja2018soft} (SAC) to train the policy with maximum entropy RL, which proved to be one of the \textit{state-of-the art} RL models. SAC encompasses two networks: the actor, representing the policy, $\pi(s; \phi)$, and the critic, $Q(s, a; \psi) = \mathbb{E}_\pi \big[R^{MaxEnt} | s, a \big]$, parameterized by $\psi$ and responsible for estimating the maximum entropy return starting on a state s and doing action a and then subsequent actions sampled according to the current policy, $\pi$. To model a probability distribution, the actor outputs a matrix of parameters of a Gaussian distribution. We use these parameters to sample a value per actuator in exploratory mode or use the mean in exploitation mode. The parameters of the critic are learned so that the model outputs the best prediction for the maximum entropy return. The actor is trained so that the input action into the critic maximizes such return. Finally, $\alpha$ is updated automatically so that the policy output in exploration mode reaches a target entropy, $\bar{H}$.

Regarding architecture, the policy comprises CNNs to process the state and sample an action with the same shape as the HDDM component. This action is then multiplied by a mask of valid actuators, and the HDDM and TT components are separated as shown in Figure \ref{fig:action_rl_unet}.  Similar to the actor, the critic employs CNNs to process the information.

\begin{figure}[h]
    \centering
    \includegraphics[width=0.55\textwidth, trim = 0cm 2cm 0cm 2cm, clip]{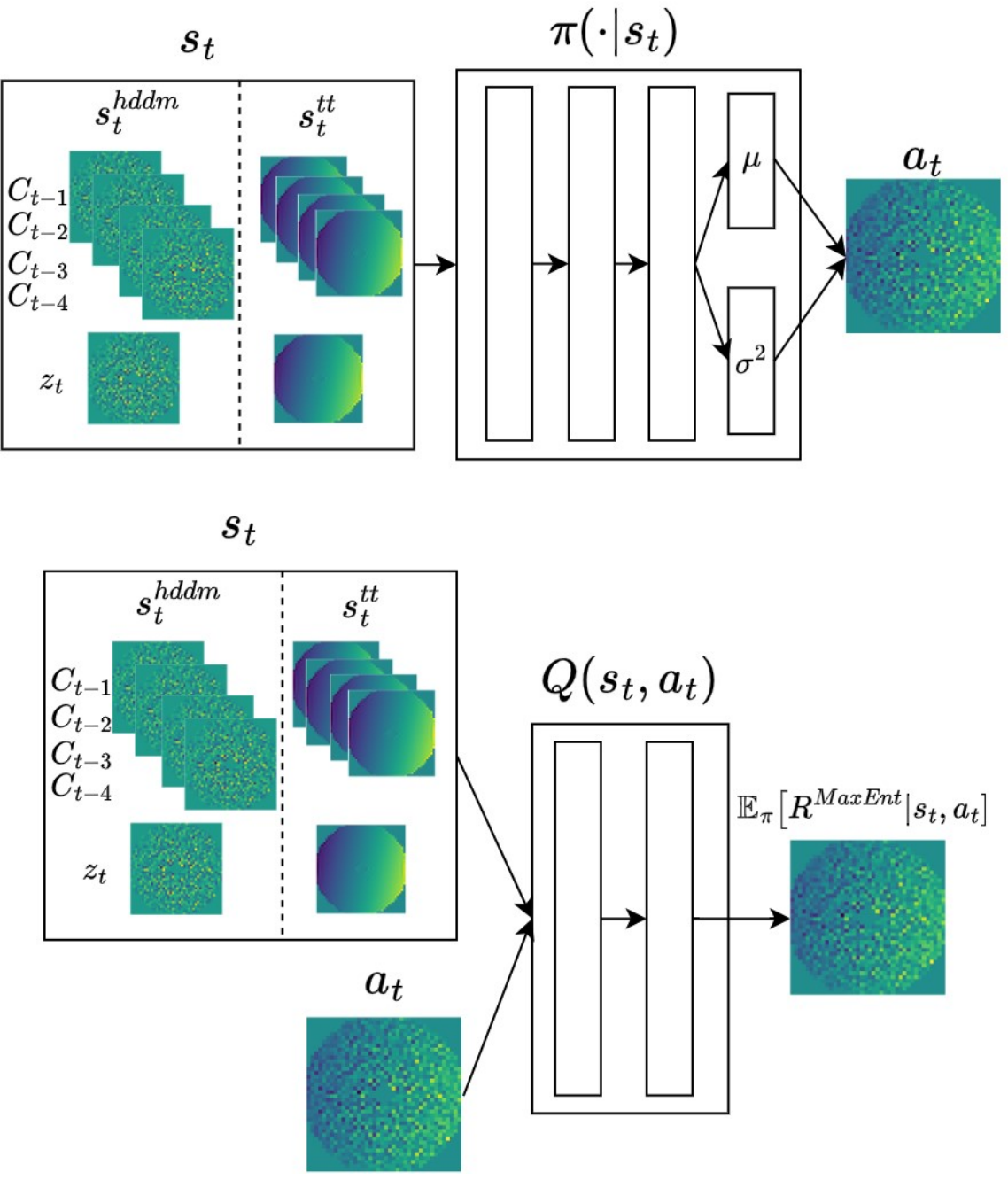}
    \caption{Top: policy which maps states to actions. Bottom: critic, which maps state and action to the estimation of maximum entropy conditioned on current state and action.}
    \label{fig:sac}
\end{figure}

Once the RL agent interacts with the AO environment during observation, data for state, actions and rewards is stored in a circular dataset known as a replay buffer with a predefined size. On update time, we sample batches of data from the replay buffer to train both the actor and the critic. SAC manages the exploration by having two channels of the output: a mean and a standard deviation. The standard deviation is used to sample small deviations on the mean for exploratory purposes and discarded once in exploitation mode. Figure \ref{fig:sac} illustrates the mentioned architecture.

\section{Experimental setting}
\label{experimental_setting}

\subsection{Simulation setting}

We use COMPASS \cite{ferreira2018numerical} to test our ML solution due to its demonstrated high performance, with the possibility to leverage multiple GPUs, and its compatibility with Python, which offers seamless integration with neural network frameworks. We aim to control a 40x40 HDDM with a supporting TT stage AO system on an 8-meter telescope. We assume a typical ExAO loop frequency of 1 kHz. The choice of 1 kHz instead of higher frequencies, as seen in many ExAO systems, is chosen because it allows for more integration time, hence decreasing the weight of noise in the error budget while increasing temporal error \cite{jovanovic2015subaru, guyon2017adaptive}. This is typically done for fainter targets where noise levels are high. In this case, predictive control becomes essential. We add a P-WFS with a resolution of 56x56 pixels for each of the four images. This translates to the ratio of P-WFS pixels and DM actuators being 1.4. This rule of thumb, typically used by system designers, leads to a more stable closed-loop behaviour compared to the typical factor of 1 derived from Fried's geometry. As we use non-linear reconstruction, we set the modulation to 0. For most of our experiments, we choose three $r_0$ values (0.08, 0.12 and 0.16) m and two P-WFS guide star magnitudes, $M$ (0 and 9). The values of $r_0$ are not tied to any specific observatory but reflect diverse atmospheric conditions, from poor to good seeing, as our method could be applied to most sites. Similarly, the values of $M$ are chosen based on diverse targets, bright and faint. The wind speed values chosen are above average on some reported sites \cite{roberts2011improved}, such as Mauna Kea,  which has installed 8-meter type telescopes. This allows us to showcase the ability of predictive control to correct for the high temporal error.

The list of the most relevant simulation parameters is available in Table \ref{tab:sim_pars}.

\begin{table}[h]
\centering
\caption{Simulation parameters}
\begin{tabular}{llll}
\hline
\multicolumn{2}{c}{\textbf{Telescope parameters}} & \multicolumn{2}{c}{\textbf{Atmospheric parameters}} \\ \hline
Diameter  (m)                      & 8               & Num. layers                    & 3                  \\
Pupil diameter (pixels)         & 256             & Altitude per layer (km)        & 0, 4, 10           \\ \cline{1-2}
\multicolumn{2}{c}{\textbf{AO loop parameters}}   & $r_0$ (m)                         & variable @ 500 nm  \\ \cline{1-2}
Loop frequency  (kHz)                & 1            & $r_0$ distribution                & 0.5/0.35/0.15      \\
Delay (frames)                  & 2               & $L_0$ (m)                         & 30                 \\ \cline{1-2}
\multicolumn{2}{c}{\textbf{Target parameters}}    & Wind speed per layer (m/s)     & 15, 15, 35         \\ \cline{1-2}
$\lambda_{target}$    ($\mu m$)              & 1.6             & Wind direction per layer (º)   & 0, 20, 180         \\ \hline
\multicolumn{2}{c}{\textbf{P-WFS parameters}}      & \multicolumn{2}{c}{\textbf{DM parameters}}          \\ \hline
Num. subapertures               & 56x56           & Correction stages                        & HDDM and TT         \\
Modulation                      & 0               & Num. actuators                 & 40x40              \\
$\lambda_{wfs}$   ($\mu m$)     & 0.85            & Modes filtered                 & 100                  \\
Noise   (e- RMS)                & 3         &                                &                   
\end{tabular}
\label{tab:sim_pars}
\end{table}

\subsection{U-Net architecture and training process}

Our U-Net architecture consists of 10 hidden layers with the typical U-Net format: an encoder part and a decoder part with skip connections bridging the encoder and decoder components. The activation function for our output layer is a hyperbolic tangent, and for the hidden layers, a leaky rectified linear unit \cite{maas2013rectifier}. Additionally, we use batch normalisation \cite{ioffe2015batch} in all layers, except for the first and last layers, to improve stability and performance.

For the training process, we gather two datasets, one for $M=0$ and another for $M=9$. Each dataset consists of 200 K data points split in 80\%/20\% training/evaluation. We sample (P-WFS, phase projected into DMs) pairs by randomly setting the actuator values on the HDDM and TT stage. For that, we sample variance values, $\sigma$, from a log uniform distribution from 0.015 to 2 $\mu m$ for the HDDM and from $7.5 \cdot 10^{-4}$ to 0.2 arcsec for the TT stage. This variance is then used to sample the actuator values from $\mathcal{N}(0, \sigma^2)$. Only to generate the dataset, we clip the maximum value for any actuator to be $\pm 5$ $\mu m$ and $\pm 0.5$ arcsec for the TT stage. Similarly to \cite{landman2020nonlinear}, the lognorm distribution shapes the data such that most phase points will have low RMS. This is key for closed-loop AO because most of the phases observed in the loop will have a low RMS. 

The input of the U-Net is the P-WFS images. Each P-WFS image is restructured into a matrix of size (4, 56, 56), where the image for each pyramid face corresponds to a channel. Due to the U-Net architecture constraints, we pad the input array, resulting in a shape of (4, 64, 64). Next, we clip the minimum value to 0 to remove any artefact coming from negative values. Through our initial testing, we discovered that with $M=9$, the U-Net's performance in a closed-loop setting was unstable, a problem we believe appears from the neural network being confused by noise, given that the issue did not appear for $M=0$. As the data from the P-WFS represents photon counts, where negative values are not physically meaningful, we opted for this specific approach to noise management which turned out to address the aforementioned problem. Finally, we normalise each input P-WFS image, $m$, by dividing by the $(max_{m} - min_{m})$ where the maximum and minimum are from all P-WFS samples in the training set.

The goal of the U-Net is to predict the phase projected into the virtual DM, which has size (40, 40), with HDDM and TT components. Because of the autoencoder nature of the U-Net, it outputs an array of size (1, 64, 64). For that, we first multiply by the mask of valid actuators, which discard values outside of the valid zone, and then we trim the edges, resulting in the shape (40, 40). The real phase projected onto the virtual DM, $c^*$, is normalised by dividing by the $max_{c^*}-min_{c^*}$, where the max and minimum are from all phase samples in the training set. During the evaluation, we multiply by a mask of valid actuators, and then we denormalise by multiplying by the same $max_{c^*}-min_{c^*}$.

We train two U-Nets, one per magnitude we experiment with (0 and 9). We use the ADAM (Adaptive Momentum) optimizer \cite{kingma2014adam} with a batch size of 32 and a starting learning rate of $2 \cdot 10^{-4}$. The weights are initialised using a Normal distribution with $\mathcal{N}(0, 0.02)$. During training, we decrease the learning rate if the evaluation set loss is not reduced after five epochs, i.e. a single pass through the whole dataset. Moreover, we use an early stopping procedure where if the evaluation loss has not decreased for 20 continuous epochs, we stop the training. The U-Net used for magnitude 0 and 9 has been trained for 131 and 60 epochs respectively.

\subsection{RL architecture and hyperparameters}

The policy is composed of three hidden convolutional layers, whereas the critic is composed of two hidden convolutional layers. Each layer has 64 filters, each with rectified linear activation functions \cite{maas2013rectifier} except for the last layers, which have no activation function in the case of the critic and have a tanh activation function in the case of the policy. For the policy, after the tanh activation function, the values are bounded between [-1, 1]. We multiply the action by a parameter, which we call the freedom parameter, to have more reasonable bounds. For the HDDM the freedom parameter is 0.1 and for the TT stage is 0.005, ending up with the bounds [-0.1, 0.1] $\mu m$ and [-0.005, 0.005] arcsec, respectively. While the action values are small, as we are integrating over time, the delta of action between steps does not need to be large. We clip the maximum value for the commands to be 10 $\mu m$. We use ADAM optimizer for the policy and critic. The weights are initialised using Xavier initialisation \cite{glorot2010understanding} for both actor and critic. Regarding the hyperparameter $\lambda$ that encourages a smooth HDDM surface, we tested a few values. We ended up choosing $10^{-4}$ for $M=0$ and $10^{-5}$ for $M=9$. Most of the RL's hyperparameters are listed in Table \ref{tab:sac-parameters}, with additional details being provided within the code repository \cite{code_for_paper}. 

As mentioned, we build the RL model atop a reconstruction, which is created by combining linear and non-linear approaches. For the RL experiments, we use the U-Net trained with 200 K training samples with optimized gains. The U-Net approach is both part of the RL model and the baseline in itself. For all RL experiments, we show the mean and standard deviation for three different seeds.

\begin{table}[h]
\centering
\caption{SAC hyperparameters}
\begin{tabular}{|l|l|l|l|}
\hline
$\boldsymbol{\gamma}$              & 0.1    & \textbf{Replay buffer size}                & $5 \cdot 10^4$ \\ \hline
\textbf{Num. filters per layer}   & 64    & \textbf{Batch size}                 & 256                   \\ \hline
\textbf{Num. hidden layers}: $\pi$ (Q) & 3 (2)      & \textbf{Optimizer} & ADAM                  \\ \hline
$\boldsymbol{\lambda}: \ M=0 \  \ (M=9)$              & $10^{-4}$  ($10^{-5}$)  & $\boldsymbol{\alpha}$                    & Automatic             \\ \hline
\textbf{Learning rate}      & $3 \cdot 10^{-4}$ &           $\boldsymbol{\bar{H}}$                  &     1600                  \\ \hline
\end{tabular}
\label{tab:sac-parameters}
\end{table}

\section{Results}
\label{Results}

\subsection{Results for non-linear reconstruction}

This section evaluates the phase reconstruction quality performed by the U-Net along the following questions:

\begin{enumerate}
    \item Does the U-Net model predictions outperform the linear reconstruction?
    \item How does it perform in closed-loop both with bright and faint stars?
    \item What is the effect of the dataset size on the quality of the U-Net model?
\end{enumerate}

\begin{figure}[h]
    \centering
    \begin{subfigure}[b]{0.48\textwidth}
        \includegraphics[width=\textwidth, trim = 0cm 0cm 0cm 0cm, clip]{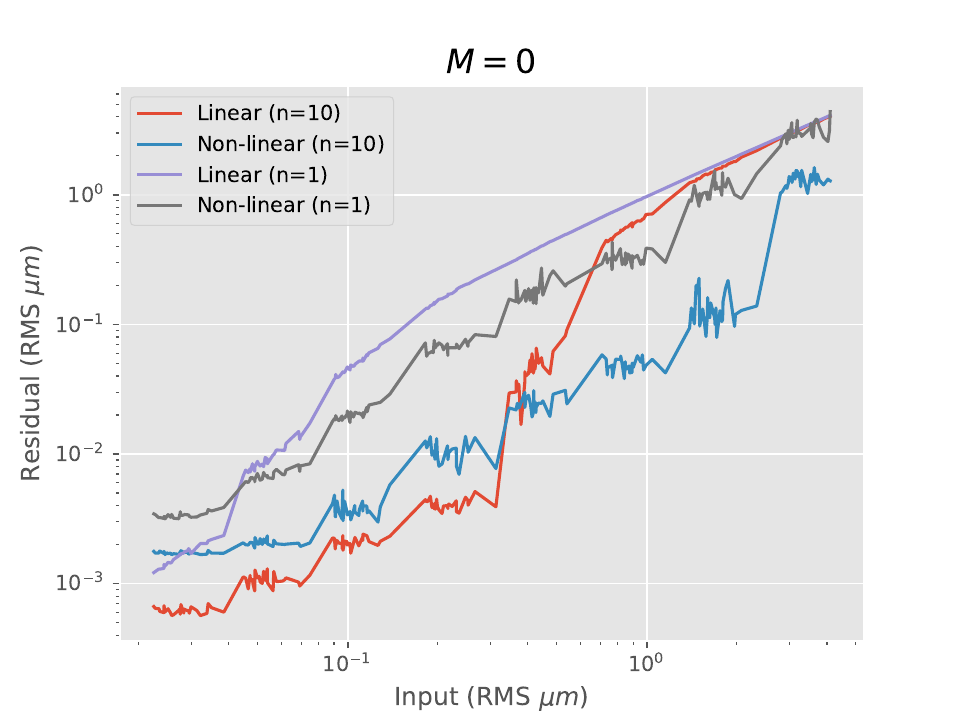}
        \subcaption{Comparison of linear and non-linear reconstruction for $M=0$.}
        \label{fig:sub1}
    \end{subfigure}
    \hfill
    \begin{subfigure}[b]{0.48\textwidth}
        \includegraphics[width=\textwidth, trim = 0cm 0cm 0cm 0cm, clip]{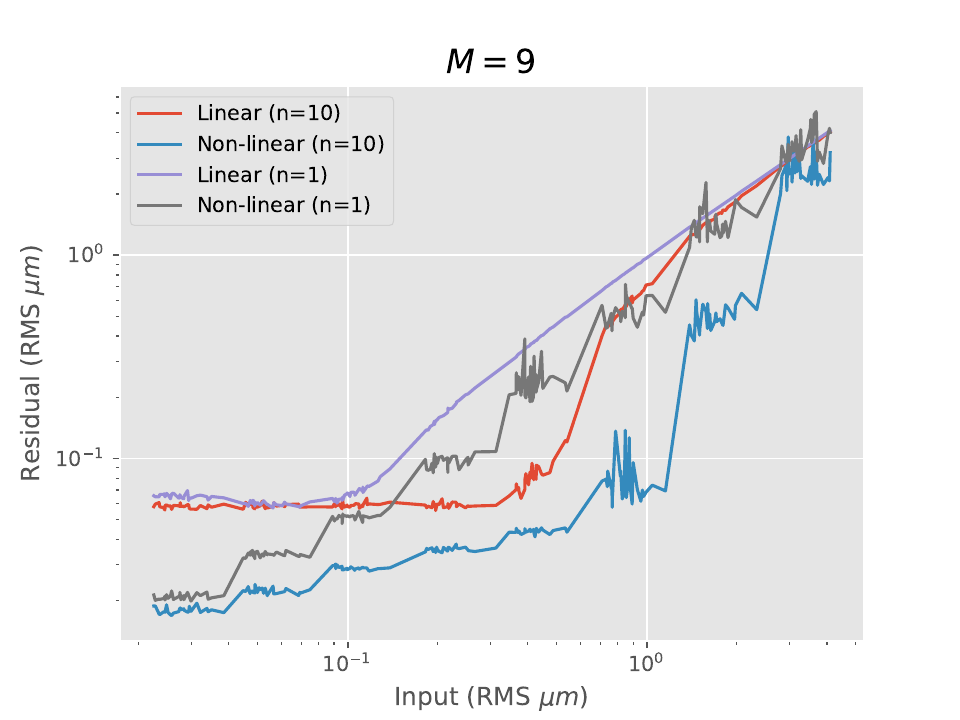}
        \subcaption{Comparison of linear and non-linear reconstruction for $M=9$.}
        \label{fig:sub2}
    \end{subfigure}
    
    \begin{subfigure}[b]{0.48\textwidth}
        \includegraphics[width=\textwidth]{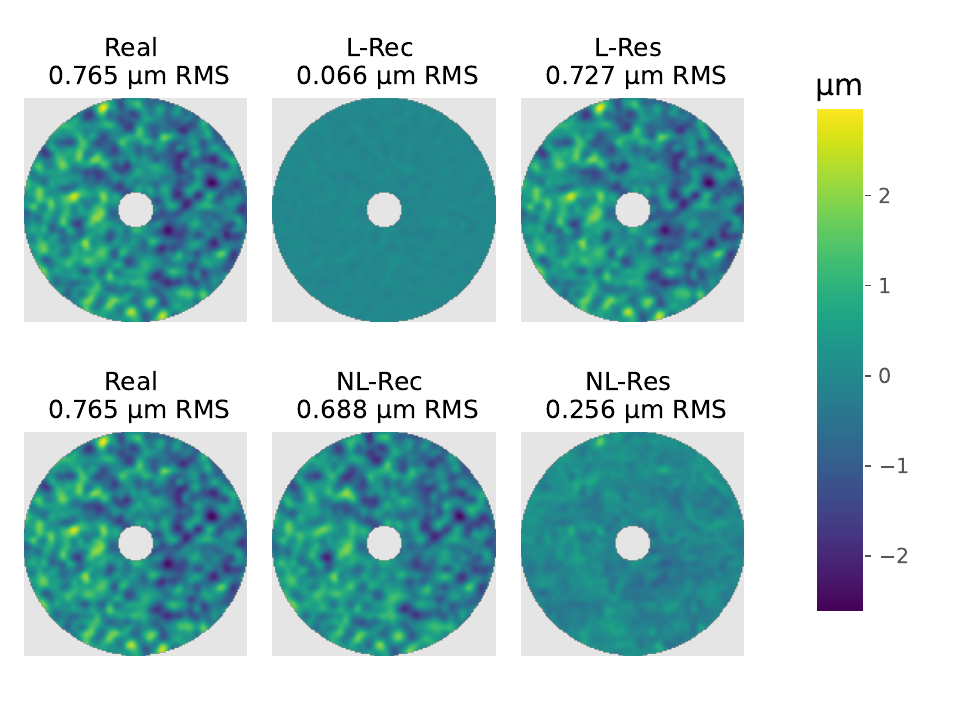}
        \subcaption{Example reconstruction for $M=0$ and n = 1.}
        \label{fig:sub3}
    \end{subfigure}
    \hfill
    \begin{subfigure}[b]{0.48\textwidth}
        \includegraphics[width=\textwidth]{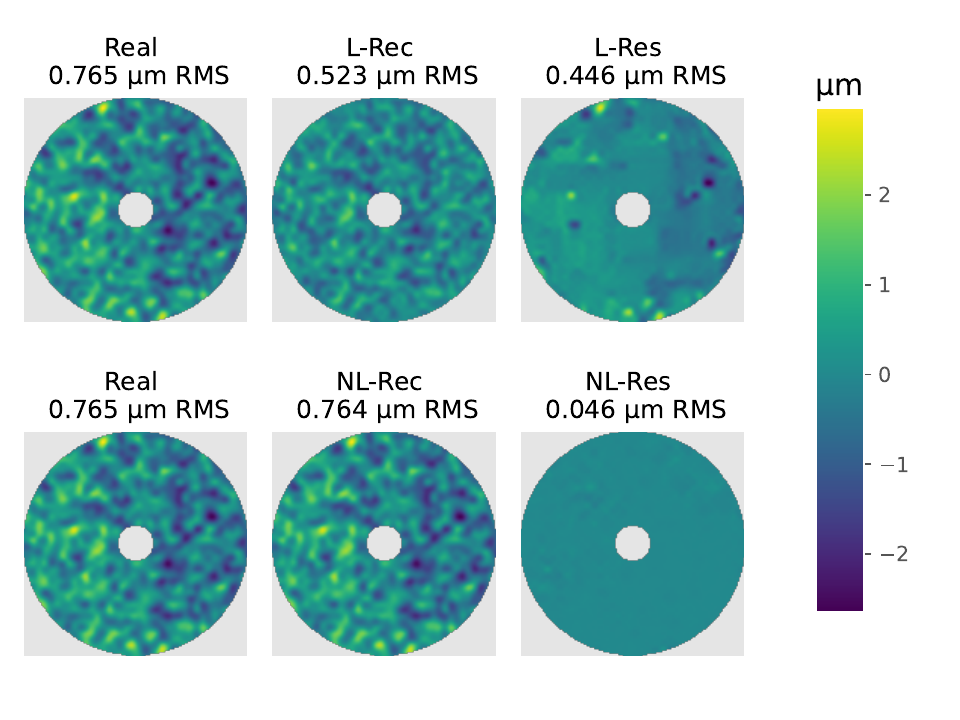}
        \subcaption{Example of reconstruction for $M=0$ and n = 10.}
        \label{fig:sub4}
    \end{subfigure}
    \caption{Assessing U-Net performance.}
    \label{fig:main_unet_fixed}
\end{figure}

We start by assessing the U-Net reconstruction accuracy. We use a range of variance levels, $\sigma^2_i$, in the range $[0.02, 2.56] \ \mu m$ for the HDDM and $[1, 128] \cdot 10^{-3}$ arcsec for the TT stage. For each variance level, we generate 30 phase screens using a Normal distribution centred at zero with a variance corresponding to our test level $\mathcal{N}(0, \sigma^2_i)$ and by moving the HDDM and the TT stage, we obtain their corresponding WFS image. Then, we estimate the phase using a reconstructor and obtain the residual root mean square (RMS) by comparing the predicted and actual phases. In addition, we treat the reconstruction problem as a closed-loop integrator with unitary gain, keeping the phase fixed. We want to observe the end result of both U-Net's and linear's predictions after making a series of 10 successive closed-loop predictions. The outcomes of these experiments are shown in Figure \ref{fig:sub1} for magnitude 0 and in Figure \ref{fig:sub2} for magnitude 9 guide star. The y-axis represents the residuals for either one prediction (n=1) or ten predictions (n=10) for both linear and non-linear reconstructions.

For n=1, the non-linear reconstruction outperforms the linear one for most tested input RMS and magnitudes except for $M=0$ and an input RMS below 0.04 $\mu m$. For n=10 and magnitude 0, the non-linear reconstruction outperforms the linear one for input RMS above 0.3 $\mu m$. With lower input errors representing a pure linear regime, the U-Net struggles to predict accurately enough. As a non-linear predictor, it tends to introduce some errors, which might make a linear reconstructor more suitable under slow turbulence and low seeing. For n=10 and magnitude 9, the non-linear one outperforms the linear one in all cases. The results indicate that there is a performance floor due to the presence of noise that the linear reconstruction cannot surpass, while the U-Net seems to be able to mitigate this effect.

From the U-Net performance plots for large input errors, we infer that it is correcting for optical gains and other non-linear effects; however, it also struggles with extremely high input RMS phases. This could be attributed to two main reasons: a) the majority of the training dataset exhibits low RMS due to the use of log-normal distribution to generate it, and b) during training, the L1 relative loss predominantly prioritises low RMS predictions. This is due to the normalization of the real phase by the L1 norm. While the relative prediction error remains similar for both small and high input RMS values, the absolute error tends to be much larger for larger RMS values. Comparing the n=1 and n=10 cases, we observe that while the residual error is continually reduced with the linear reconstruction as the number of iterations increases, the U-Net does it much faster.

To provide a deeper understanding, we also present specific examples of reconstructions in the subfigures of Figure \ref{fig:sub3} and \ref{fig:sub4}, including linear reconstruction (L-Rec), non-linear reconstruction (NL-Rec) and their correspondent residuals (Res). In Figure \ref{fig:sub3}, we show the case of a single-step reconstruction (n=1) under high input RMS and magnitude 0. The non-linear approach performs significantly better at reducing the residual variance than the linear reconstruction, which offers a limited correction. Figure \ref{fig:sub4} showcases the outcomes after ten iterations of reconstruction in the same case. The U-Net provides an excellent reconstruction result, whereas the linear method still exhibits inconsistencies and mispredictions.

\begin{figure}[h]
    \centering
    \includegraphics[width=0.9\textwidth, trim = 0cm 1cm 0cm 0.5cm, clip]{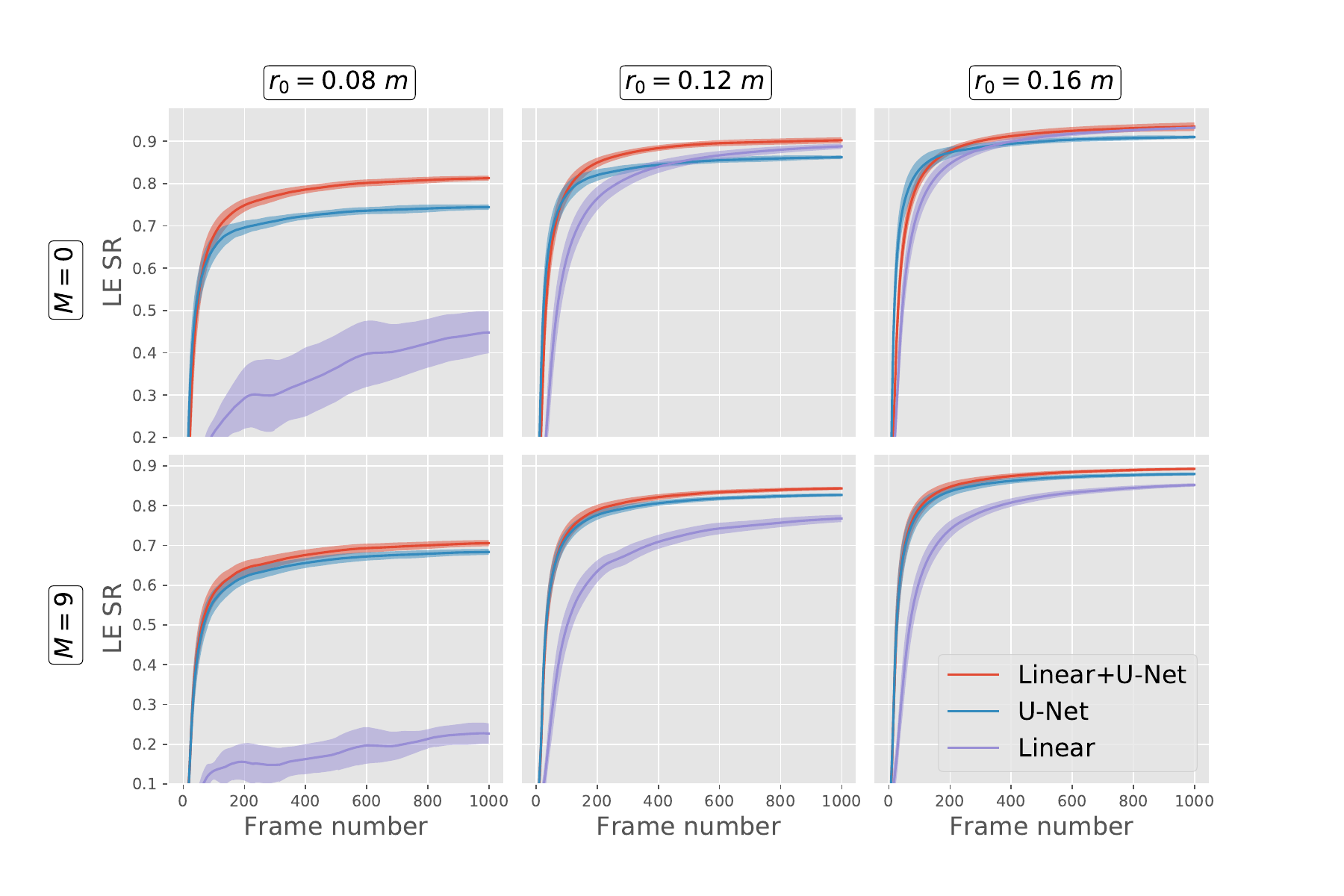}
    \caption{Closed-loop performance of linear, non-linear (labelled U-Net) and combination of linear and non-linear (labelled Linear+U-Net) reconstructions for different atmospheric conditions and different guide star values for the P-WFS. We show the mean and standard deviation of the result over five seeds of long exposure Strehl Ratio (LE SR) in the y-axis.}
    \label{fig:closed_loop_unet}
\end{figure}

For the second question, we want to assess the U-Net performance under realistic atmospheric turbulence in a closed-loop scenario. For that, we test an integrator with three different reconstruction methods: linear, U-Net's non-linear reconstruction and a combination of linear and non-linear reconstructions. As an initial step, we fine-tuned the gains for all controller types and for each simulation scenario. The gain optimization was performed by testing the integrator over a range of gains in closed-loop operation over 1000 frames and obtaining the gains that produced the best long exposure Strehl Ratio (LE SR). Figure \ref{fig:closed_loop_unet} shows the closed-loop results in terms of LE SR for the three distinct reconstruction methods. These tests spanned various atmospheric conditions, characterised by values of $r_0$ (0.08, 0.12, 0.16) m, and were executed under two different guide star magnitudes ($M=0$ and $M=9$). For the case of $M=9$ and the linear reconstruction controller, we experimented with using the same clipping strategy as the one used for the U-Net without any improvement.

\begin{figure}[h]
    \centering
    \includegraphics[width=0.85\textwidth, trim = 0cm 0.5cm 0cm 0.5cm, clip]{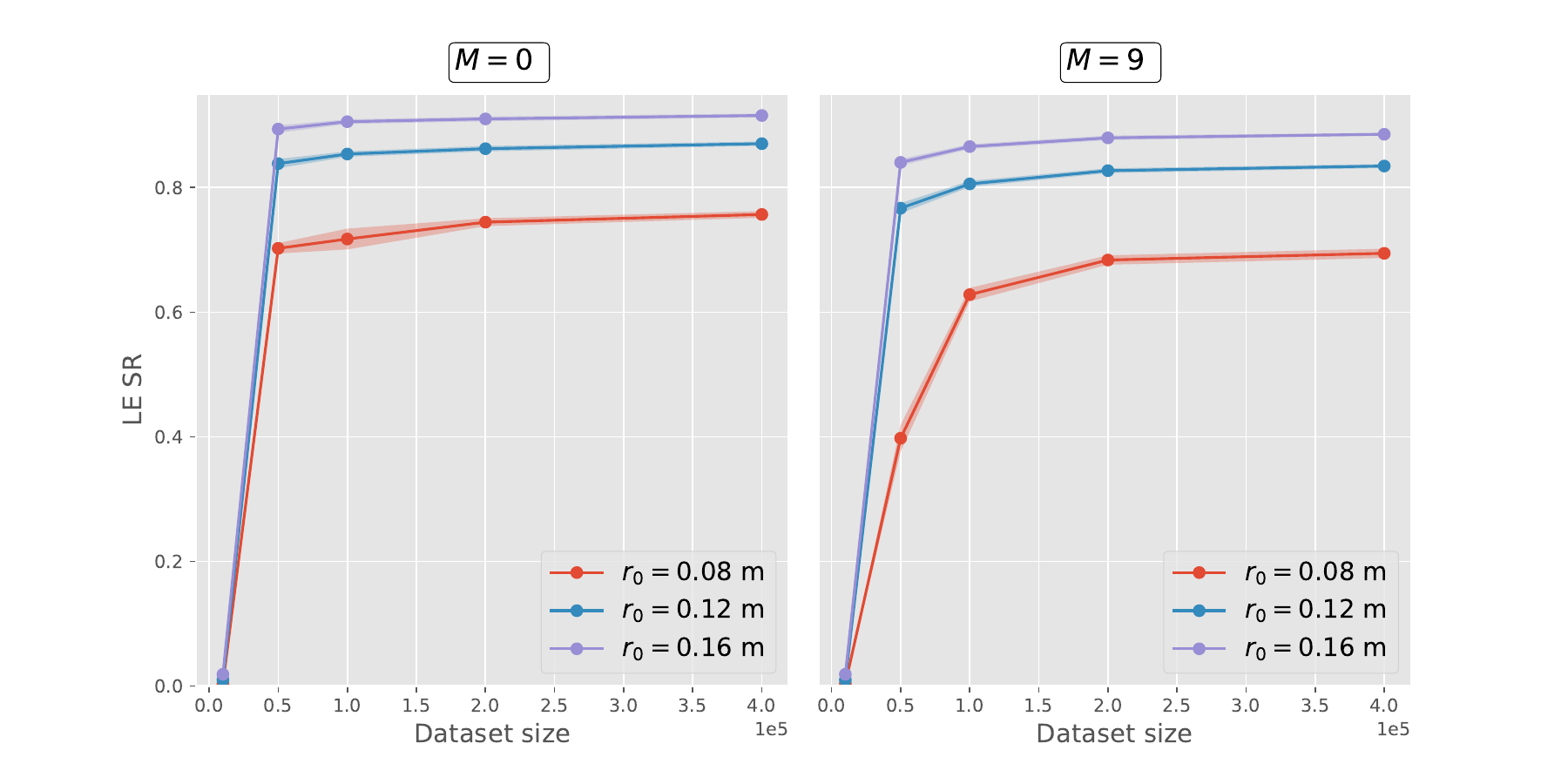}
    \caption{Comparison of results based on dataset size. We show the mean and standard deviation of the result over five seeds.}
    \label{fig:dataset_difference}
\end{figure}

The Figure shows that on the most favourable conditions ($r_0 = (0.12, 0.16) \ m$ and $M = 0$), the U-Net alone is not able to match the linear reconstructor performance. Our hypothesis is that in the cases of mild to low turbulence, the P-WFS response lies mostly within the linear regime (good linear reconstructor performance) and the U-Net accuracy in such a regime, although quite good, appears limited. Instead, we need to combine the non-linear and linear reconstruction to achieve better or similar results. For all other scenarios, the U-Net surpasses the linear approach. Notably, at $r_0=0.08$ m, the linear approach struggles to close the loop, while the U-Net alone succeeds. Interestingly, the U-Net is able to manage noise much better than the linear controller in this simple form, in which no noise mitigation strategy was implemented (e.g. thresholding). We recall here that, for linear reconstruction, the P-WFS pixel data is not processed through any means and refers the reader to equations \ref{eq:1_interaction}, \ref{eq:2_reconstruction} and \ref{eq:3_integration} as of how the P-WFS response is used.  Such native robustness against noise must be inherited from the autoencoder nature of the U-Net, which compresses and decompresses information, therefore being able to partially filter out the noise. In the work of \cite{smith2023study}, authors have conducted research on a similar U-Net architecture with a Shack-Hartmann WFS, demonstrating enhanced performance compared to the linear reconstruction in a noisy setting. Finally, the combination of both linear and non-linear reconstructions appears to be the one that outperforms the linear approach in all cases, demonstrating the benefits over conventional linear reconstruction.

Finally, for the third question, we want to estimate the effect of the dataset size on performance. For that, we train the network with the following dataset lengths (10 K, 50 K, 100 K, 200 K, 400 K) for $M=0$ and $M=9$. Figure,  \ref{fig:dataset_difference} shows the closed loop final performance over 1000 frames (same experiment as Figure \ref{fig:closed_loop_unet}) for the different amounts of data and values of $r_0$. We observe different requirements in terms of training dataset size for each magnitude. For $M=0$ and all $r_0$ values, after 50 K samples, the performance increments start to be small. For $M=9$, a larger dataset seems to be required under stronger turbulence. For both magnitudes and all $r_0$, 200 K samples appear to be the sweet spot between increasing accuracy and collecting more data.

\subsection{Results for predictive control}

This section evaluates the RL predictive control along the following questions:

\begin{enumerate}
    \item Does the RL predictive control increase the performance over a baseline integrator with a mixed linear and non-linear reconstruction with optimized gain both with a bright and with a faint guide star?
    \item Is the RL model performing predictive control?
    \item Is the RL model able to control the TT stage given the defined model?
    \item How is the RL model able to adapt to changing atmospheric conditions?
\end{enumerate}

\begin{figure}[h]
    \centering
    \includegraphics[width=0.85\textwidth, trim = 0cm 1cm 0cm 0.5cm, clip]{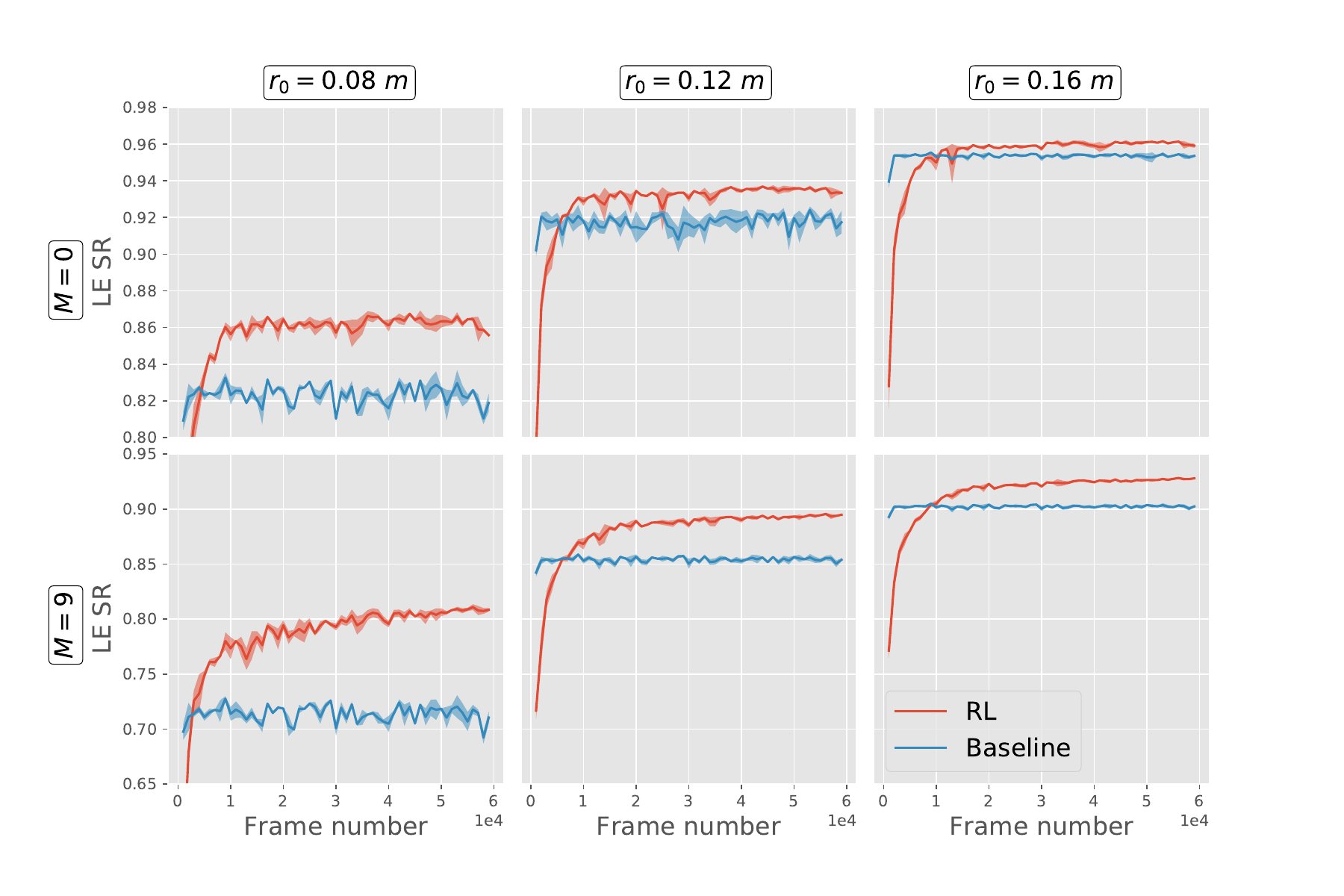}
    \caption{Result of training the RL controller compared with the baseline integrator with a mixed linear and non-linear reconstruction.}
    \label{fig:closed_loop_rl}
\end{figure}

Regarding the first question, we examine the RL model performance for the different values of $r_0$ and star brightness. We train the RL predictive control model for 60 K frames, measuring the SR in exposures of a thousand frames. For 30 K frames, the agent is in exploration mode. After 30 K frames, we switch to exploitation, and the action is just the mean. Figure \ref{fig:closed_loop_rl} shows the results of the RL controller against our baseline, the integrator, with a combination of linear and non-linear reconstructions. We do not show the integrator with linear reconstruction only because the performance is equal to or lower than the one obtained with the combination).

We can see that the RL model outperforms the baseline integrator in all cases after learning for less than 10 K frames. The algorithm shines the most when looking at $M=9$. After 60 K frames, we observe the RL outperforms the integrator by as much as 10 points in the lowest $r_0$ and over 2 points in the highest $r_0$. For a bright star, $M=0$, the improvements are still significant, around 4 points for the lowest $r_0$ and 1 point for the highest $r_0$. This behaviour is what we expected, as it is well known that the temporal error increases in the case of a fainter star because the integrator gain must be reduced to mitigate the propagation of the noise error.

By comparing the Btt mode values of the RL controller and those of an ideal baseline with zero delay, we can address questions two and three: whether the RL controller is performing predictive control and whether it is correcting the TT error. For this test, we choose the simulation with the least error from other sources ($M=0$, $r_0=0.16 \ m$).

\begin{figure}[h]
    \centering
    \includegraphics[width=0.85\textwidth, trim = 0cm 1cm 0cm 0.5cm, clip]{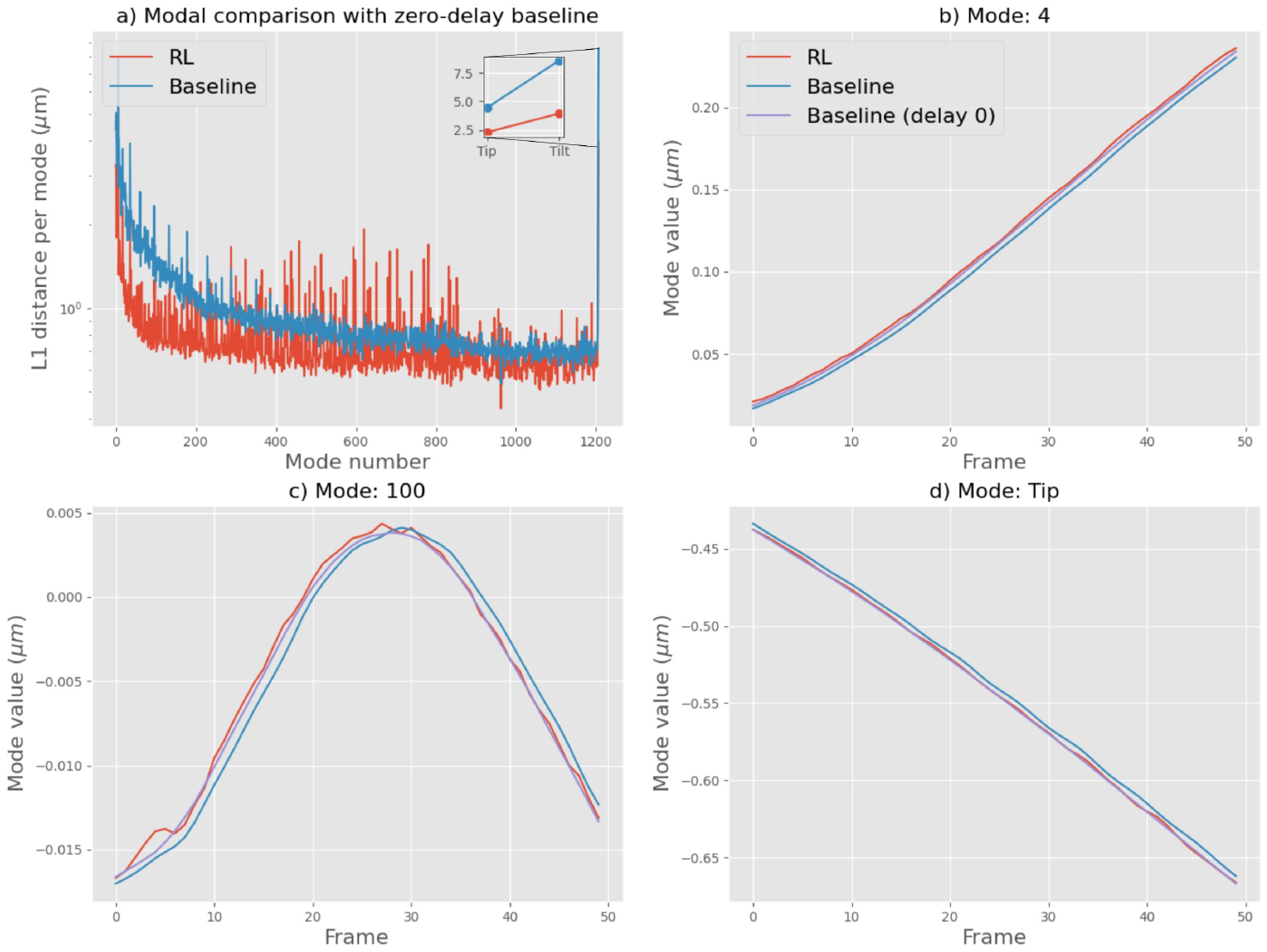}
    \caption{Predictive control and TT error correction test with modal decomposition. Figure (a) shows the sum of L1 distance per mode. Figures (b), (c) and (d) show the predictive control capabilities for modes 4, 100 and Tip for 50 frames.}
    \label{fig:modal_decomp}
\end{figure}

Figure \ref{fig:modal_decomp} (a) shows the sum of the L1 distance of each mode from a baseline integrator with zero delay with each mode from a baseline and RL with two frames of delay over 2 K frames. This can be represented mathematically as $dist(i)=\sum_{t=0}^{t=2000} |mode_{i,t}^{d0}-mode_{i,t}^{controller}|$ where $i$ is the index for a mode, $d0$ indicates baseline with delay 0 and controller indicates either baseline with delay 2 or RL with delay 2. For this test, the RL controller was previously trained for 60 K frames. As we can observe, on average, the RL controller is closer to the baseline without delay, hinting that the RL controller is performing predictive control. The last two modes are the TT modes, and for clarity, we provide a zoomed-in plot. In this zoomed-in plot, we can observe that the RL controller has better metrics also with the TT modes.

Figures \ref{fig:modal_decomp} (b), (c), and (d) compare the mode values for 50 frames after 950 iterations of simulation. The plots show three controllers: RL, baseline and baseline delay 0 for modes 4, 100 and Tip. As shown, the RL learns to be closer (albeit not perfectly) to the baseline with delay 0 in all the cases, finally confirming that the RL controller is a predictive controller and can correct for TT error.

\begin{figure}[h]
    \centering
    \includegraphics[width=0.8\textwidth, trim = 0cm 1cm 0cm 0.5cm, clip]{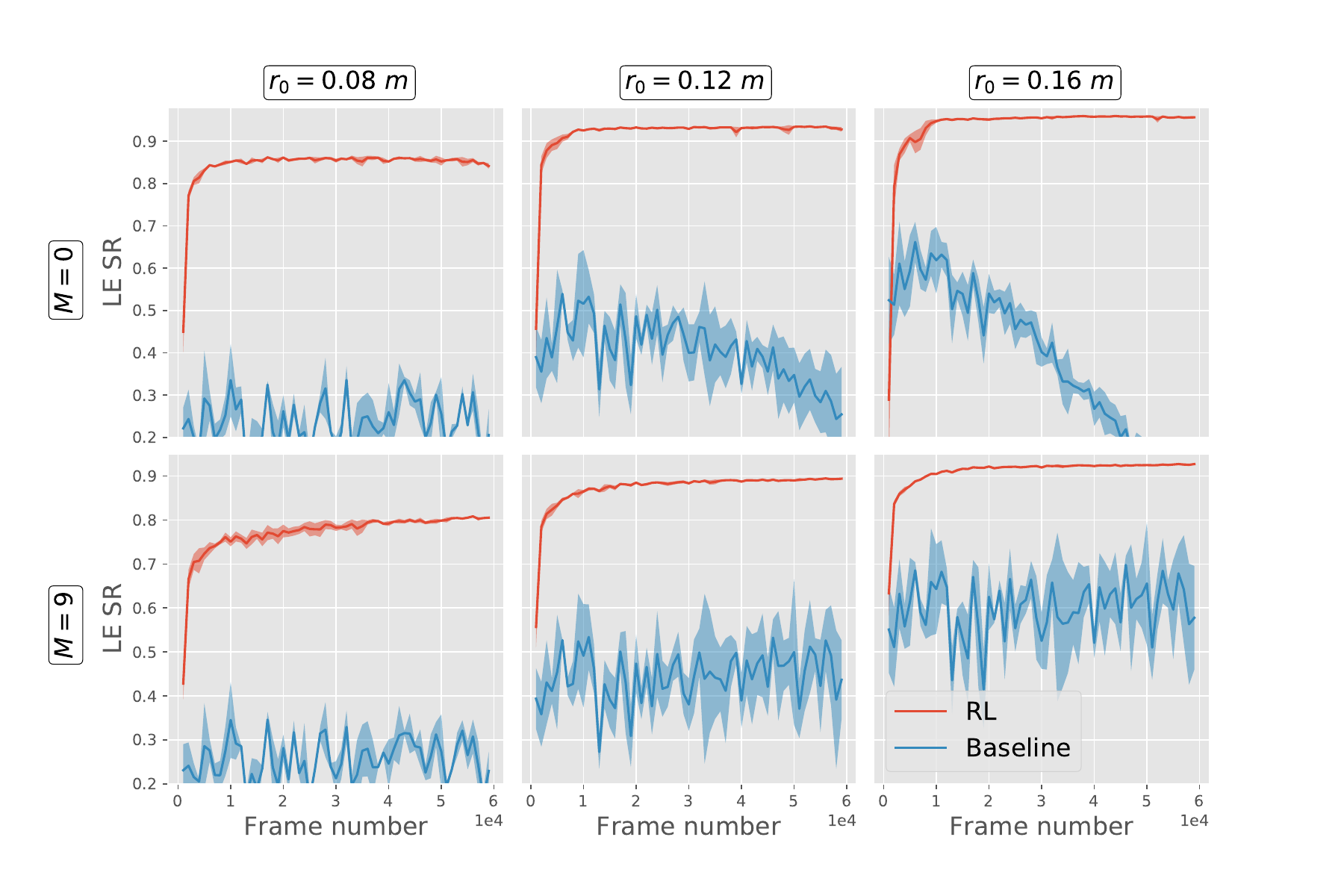}
    \caption{Result of training the RL controller compared with the baseline integrator with a mixed linear and non-linear reconstruction where the tip-tilt gain is reduced by 10.}
    \label{fig:closed_loop_rl_tt}
\end{figure}

To evaluate if RL can help to correct for a large TT error, we prepared another experiment where we reduced the tip-tilt gain to see if the RL model can compensate for that. Figure \ref{fig:closed_loop_rl_tt} shows such an experiment in which the gain for the TT stage was reduced by a factor of 10. Indeed, the RL agent is able to mitigate the TT error that appears due to the decreased TT gain. In all experiments, the baseline does not provide acceptable performance, and the RL model learns how to solve the reduced TT gain issue.

\begin{figure}[h]
    \centering
    \includegraphics[width=0.85\textwidth, trim = 1cm 2cm 1cm 2cm, clip]{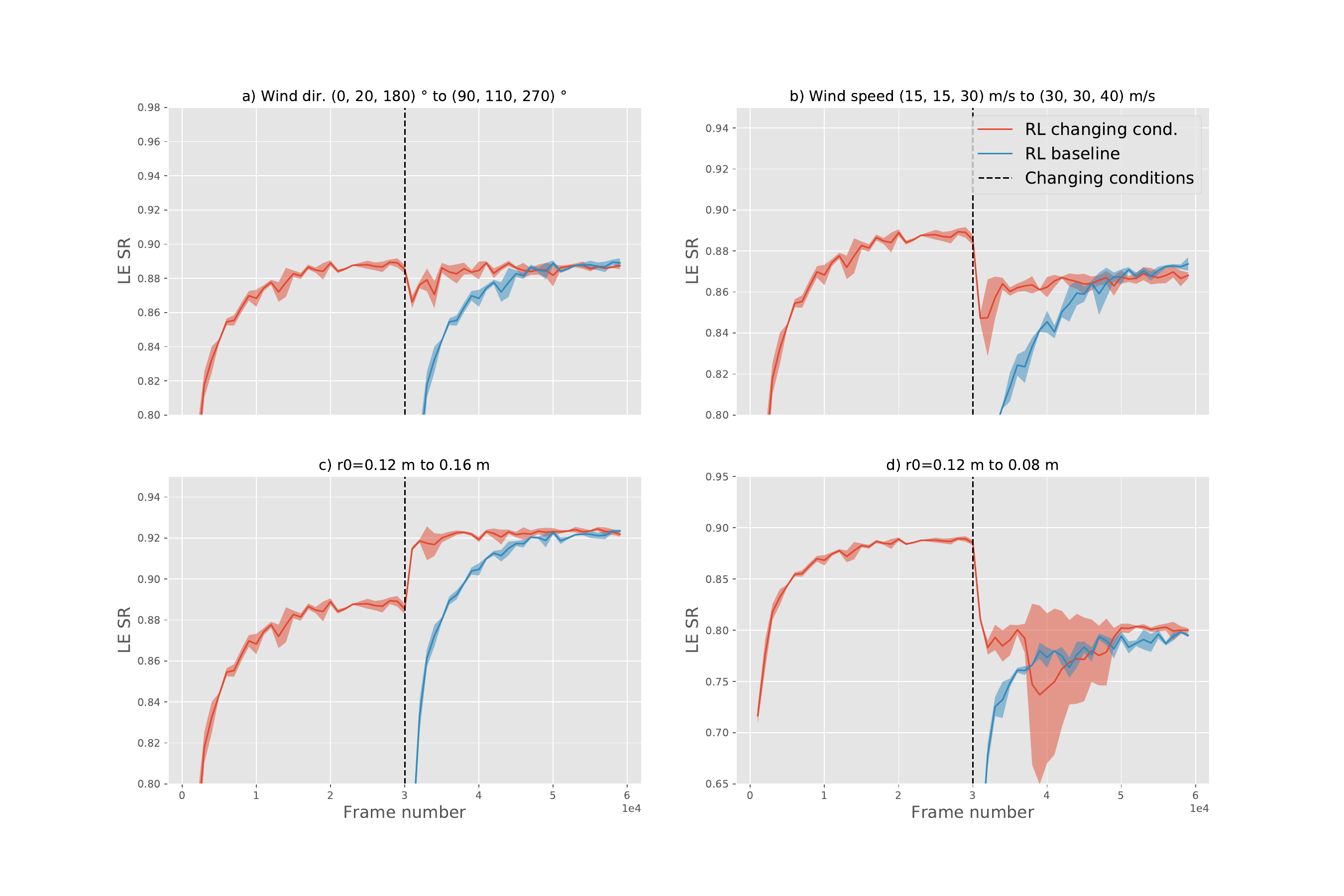}
    \caption{Changing conditions experiments.}
    \label{fig:changing_conditions_RL}
\end{figure}

Finally, we address the last question by doing an additional experiment: once the RL agent is trained for 30 K frames, we radically change the atmospheric conditions. We select one of our previous experiments, $M=9$ and $r_0=0.12$ m, and we test four cases of changing conditions: (a) change the direction of each atmospheric layer by 90 degrees, (b) change the wind speed from (15, 15, 30) m/s to (30, 30, 40) m/s, and (c) change the $r_0$ from 0.12 m to 0.16 m in one case and (d) to 0.08 m in another case. In this experiment, we are constantly exploring and reset the replay buffer when the atmospheric conditions are changed to avoid contamination by previous atmospheric conditions. Figure \ref{fig:changing_conditions_RL} shows the result of such an experiment in which we compared the always exploring RL agent described above against another RL controller that starts training from scratch in the new conditions. The baseline RL controller starts training at the frame after changing conditions. We want to check that adjusting to new conditions is faster than training from scratch. For all experiments, the RL controller appears to adapt faster than if one would train it from scratch. For one seed of experiment (d), the performance crashed for 10 K frames but was able to get back on track. As the changes are radical, the stability is expected to be challenged.

\subsection{Timing results}

\subsubsection{Critical path time analysis}

This section analyses the critical path of our ML system, i.e. the time it takes for the controller to execute the operations required to map the WFS image to commands. Concretely, the sum of the inference time of the U-Net model for phase reconstruction, and the RL policy for predictive control correction. All experiments are conducted on a single RTX 4080 (16 GB) and an Intel Core i7-12700 processor. For this evaluation, we incorporate our models into NVIDIA's TensorRT framework \cite{nvidia_tensorrt} for accelerated neural network inference. We measure 9000 inference times for each model following an initial 1000 inferences for warm-up. The analysis does not consider the time it takes to move data from CPU to GPU, since it has been shown by other authors \cite{plante2022high} that data from WFS can be efficiently transported from the acquisition interface to GPU memory, or additional operations such as integration.

\begin{figure}[h]
    \centering
    \includegraphics[width=0.6\textwidth, trim = 0cm 0cm 0cm 1cm, clip]{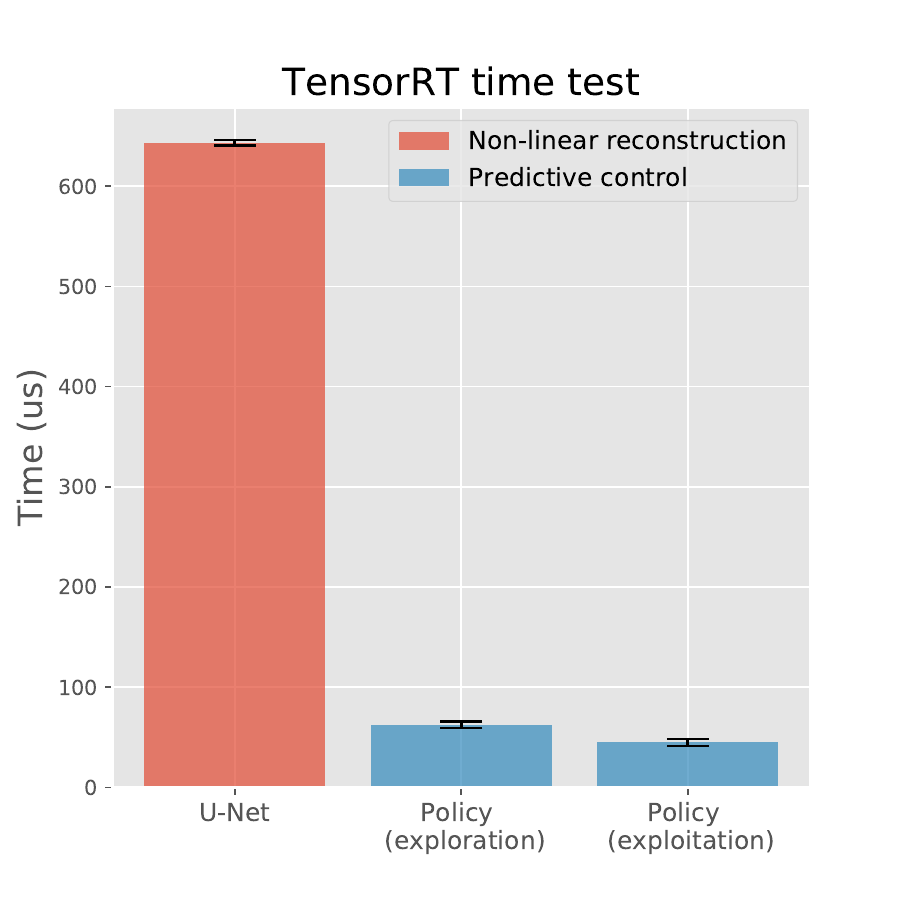}
    \caption{Inference time results for our models in TensorRT.}
    \label{fig:trt_times}
\end{figure}

Figure \ref{fig:trt_times} shows the mean and standard deviation inference times for the U-Net (labelled as U-Net) and the RL policy considering the two execution modes: exploration and exploitation (labelled as Policy (exploration) and Policy (exploitation) respectively) described in Section \ref{Methods}.  As seen in the Figure, the sum of the two operations, U-Net and RL policy, remains below 1 ms, complying with the real-time requirement of an ExAO system running at a loop frequency of 1 kHz. It is important to remark that our ML system also includes the training time of the RL agent. However, it does not form part of the critical path as the training can be conducted in parallel to the inference.

\subsubsection{Training time analysis}

The ExAO system has real-time requirements to guarantee that the evolution of the atmosphere is captured by the rate of adaptation. Not fulfilling it would imply a performance degradation of the AO system. We track the time per update given the batch size of 256 for 9000 updates after a warm-up period of 1000 updates during the online training of the RL model. The training code is implemented in Python with a Pytorch backend. We observe time per update of $72 \pm 1$ ms. In Figure \ref{fig:closed_loop_rl} from the previous section, we surpass the integrator's performance at around 10 K frames at maximum. Considering we update once per frame and the update time previously mentioned, it takes about 12 minutes to surpass the integrator performance. Ideally, the training would process the data as fast as the inference. While the training times currently do not match the efficiency of the inference times, there is significant potential for improvement. First, we could start with a policy that has been pretrained under a set of atmospheric conditions in simulation, then, the training would primarily consist of a fine-tuning task which could potentially speed up the overall training process. Second, we could code the training pipeline in C++. Moreover, we are exploring a few ideas, such as using large batch sizes and separating the training on different GPUs, training different policies at the same time and mixing the weights at certain points in time, using low precision training \cite{bjorck2021low} or sharing some layers in the actor and critic for training in fewer number of samples \cite{cobbe2021phasic}.

\section{Conclusions and Future Work}
\label{conclusions_future_work}

In this paper, we have introduced a ML solution for ExAO control for a system equipped with a high-order P-WFS with no modulation used to drive two correction stages: HDDM and TT. We demonstrated how combining a U-Net deep neural network for phase reconstruction and a model-free RL agent for predictive control can effectively address non-linearities that affect AO control. The U-Net, combined with a linear reconstruction, improves the performance of a classical integrator, especially for simulation settings with a low $r_0$ and faint guide stars. The RL significantly outperforms the baseline integrator, especially in adverse atmospheric conditions and with a dim star. Moreover, we have shown the robustness of our ML solution to changing atmospheric conditions. Finally, we have shown that the inference operation of the involved deep neural networks is within the critical path of a 1 kHz ExAO system, ensuring its correct operation. 

We want to highlight that although our simulations show promising results, COMPASS computes the atmospheric evolution under the assumption of frozen flow, simplifying the predictive control problem. Other methods have been tested on telemetry data, e.g.\cite{wong2021predictive, chen2021performance, hafeez2022forecasting}, which have extra complexity of non-frozen flow statistics, however in our case, as we are using trial and error through RL, we are limited to test on simulation, the bench or on-sky. As we want to deploy on 8-meter or more class telescopes which hold the secondary mirror with spiders, one possible hindrance is petal modes caused by various factors such as the low-wind effect \cite{milli2018low}. However, we note that it has been shown that unmodulated pyramids can sense petals \cite{engler2019pyramid}. Moreover, even if petals would be a problem for a specific telescope, we could modify the modal basis to interpolate the actuators in the petal-affected zones \cite{bertrou2020petalometry}. Furthermore, we want to mention that the instance used of linear reconstruction is built simply by taking the generalised inverse on the interaction matrix and using a global gain for all actuators. Some improvements could be used as modal gain optimization to provide a better baseline against our results \cite{gendron1994astronomical}. Finally, we discuss the goal of the RL predictive controller, which is not to provide a theoretically optimal predictive controller. It is to build a predictive controller which is self-adaptable to ongoing conditions and compensates for a fraction of temporal and user frequent maintenance solving a corresponding engineering problem in AO operation.

As for future work, we propose different paths. As regards the U-Net, there is room for improvement to predict the phase accurately for any amplitude. As seen in Figure \ref{fig:main_unet_fixed}, this is not the case, as for larger phase amplitudes, while better than the linear reconstruction, it still partially fails. In order to address this issue, exploring other architectures, other loss functions or ways to generate the dataset might be important (for example, generating the dataset taking into account a different modal basis). Moreover, it would be interesting to test the framework described in this paper in the case of a woofer-tweeter DM configuration. As we are controlling two correction stages already (HDDM and TT), we believe that we should be able to apply the same kind of approach without any major modification. Another interesting new topic would be to mix the reward function with the science camera information, such as the measured SR or any other particular feature in the image. The operation time of our ML solution to guarantee the required timing of the AO system is a critical aspect as well. For the training, we are investigating ideas such as deploying pretrained policies and fine-tuning, low-precision training, multiple GPU training, sharing policy and critic networks or implementing the code fully in C++. Another path forward could be to investigate whether linear or non-linear models are best suited for predictive control with reinforcement learning.

Finally, the ultimate step would be to test these methods on-sky. We are currently implementing our methods in Subaru's SCExAO instrument \cite{jovanovic2015subaru} and in a bench at Observatoire de Paris simulating Sphere+ \cite{goulas2023saxo+}. Both systems are ExAO systems within an 8-meter telescope. This implementation will involve two processes, one for training and one for inference and sharing data between the two. Once the results are confirmed on the bench in real-time, the first on-sky tests will start to roll out.

\section*{Funding}
This project has received funding from the European Union's Horizon 2020 research and innovation program under the Marie Sklodowska-Curie grant agreement No 873120.

\section*{Acknowledgments}
We want to thank Jesse Cranney and Charles Gretton for their valuable insights and constructive discussions throughout the course of this project. We also would like to thank Eric Gendron for his valuable input during parts of this project. GPT-4 and Grammarly have been used for grammatical corrections.

\bibliographystyle{unsrt}  
\bibliography{references}

\begin{thebibliography}{10}

\bibitem{guyon2018extreme}
Olivier Guyon.
\newblock Extreme adaptive optics.
\newblock {\em Annual Review of Astronomy and Astrophysics}, 56:315--355, 2018.

\bibitem{ragazzoni1996pupil}
Roberto Ragazzoni.
\newblock Pupil plane wavefront sensing with an oscillating prism.
\newblock {\em Journal of modern optics}, 43(2):289--293, 1996.

\bibitem{deo2021correlation}
Vincent Deo, {\'E}ric Gendron, Fabrice Vidal, et~al.
\newblock A correlation-locking adaptive filtering technique for minimum variance integral control in adaptive optics.
\newblock {\em arXiv preprint arXiv:2103.09921}, 2021.

\bibitem{stahl1995optimization}
Steven~M Stahl and David~G Sandler.
\newblock Optimization and performance of adaptive optics for imaging extrasolar planets.
\newblock {\em The Astrophysical Journal}, 454(2):L153, 1995.

\bibitem{guyon2017adaptive}
Olivier Guyon and Jared Males.
\newblock Adaptive optics predictive control with empirical orthogonal functions (eofs).
\newblock {\em arXiv preprint arXiv:1707.00570}, 2017.

\bibitem{males2018ground}
Jared~R Males and Olivier Guyon.
\newblock Ground-based adaptive optics coronagraphic performance under closed-loop predictive control.
\newblock {\em Journal of Astronomical Telescopes, Instruments, and Systems}, 4(1):019001--019001, 2018.

\bibitem{guo2006wavefront}
Hong Guo, Nina Korablinova, Qiushi Ren, et~al.
\newblock Wavefront reconstruction with artificial neural networks.
\newblock {\em Optics Express}, 14(14):6456--6462, 2006.

\bibitem{osborn2012using}
James Osborn, Francisco Javier De~Cos Juez, Dani Guzman, Timothy Butterley, Richard Myers, Andr{\'e}s Guesalaga, and Jesus Laine.
\newblock Using artificial neural networks for open-loop tomography.
\newblock {\em Optics express}, 20(3):2420--2434, 2012.

\bibitem{swanson2018wavefront}
Robin Swanson, Masen Lamb, Carlos Correia, et~al.
\newblock Wavefront reconstruction and prediction with convolutional neural networks.
\newblock In {\em Adaptive Optics Systems VI}, volume 10703, page 107031F. International Society for Optics and Photonics, 2018.

\bibitem{suarez2018improving}
Sergio~Luis Su{\'a}rez~G{\'o}mez, Carlos Gonz{\'a}lez-Guti{\'e}rrez, Enrique D{\'\i}ez~Alonso, Jes{\'u}s~Daniel Santos~Rodr{\'\i}guez, Maria~Luisa S{\'a}nchez~Rodr{\'\i}guez, Jorge Carballido~Landeira, Alastair Basden, and James Osborn.
\newblock Improving adaptive optics reconstructions with a deep learning approach.
\newblock In {\em Hybrid Artificial Intelligent Systems: 13th International Conference, HAIS 2018, Oviedo, Spain, June 20-22, 2018, Proceedings 13}, pages 74--83. Springer, 2018.

\bibitem{landman2020nonlinear}
Rico Landman and SY~Haffert.
\newblock Nonlinear wavefront reconstruction with convolutional neural networks for fourier-based wavefront sensors.
\newblock {\em Optics Express}, 28(11):16644--16657, 2020.

\bibitem{wong2023nonlinear}
Alison~P Wong, Barnaby~RM Norris, Vincent Deo, Peter~G Tuthill, Richard Scalzo, David Sweeney, Kyohoon Ahn, Julien Lozi, S{\'e}bastien Vievard, and Olivier Guyon.
\newblock Nonlinear wave front reconstruction from a pyramid sensor using neural networks.
\newblock {\em Publications of the Astronomical Society of the Pacific}, 135(1053):114501, 2023.

\bibitem{landman2024making}
Rico Landman, Sebastiaan Haffert, Jared Males, Laird Close, Warren Foster, Kyle Van~Gorkom, Olivier Guyon, Alex Hedglen, Maggie Kautz, Jay Kueny, et~al.
\newblock Making the unmodulated pyramid wavefront sensor smart. closed-loop demonstration of neural network wavefront reconstruction with magao-x.
\newblock {\em arXiv preprint arXiv:2401.16325}, 2024.

\bibitem{smith2022enhanced}
Jeffrey Smith, Jesse Cranney, Charles Gretton, et~al.
\newblock Enhanced adaptive optics control with image to image translation.
\newblock In {\em The 38th Conference on Uncertainty in Artificial Intelligence}, 2022.

\bibitem{jorgenson1994wavefront}
MB~Jorgenson and GJM Aitken.
\newblock Wavefront prediction for adaptive optics.
\newblock In {\em Active and adaptive optics: ESO Conference and Workshop Proceedings, Proceedings of the ICO-16 (International Commission for Optics) Satellite Conference on Active and adaptive optics, held August 2-5, 1993, Garching near Munich, Germany (held in conjunction with the 16th Congress of the International Commission for Optics, Budapest, Hungary, August 9-13, 1993). Edited by Fritz Merkle. Published by European Southern Observatory, Garching near Munich, 1994, p. 143}, volume~48, page 143, 1994.

\bibitem{lloyd1996spatio}
Michael Lloyd-Hart and Patrick McGuire.
\newblock Spatio-temporal prediction for adaptive optics wavefront reconstructors.
\newblock In {\em Adaptive Optics. ESO Conference and Workshop Proceedings, Proceedings of a topical meeting, held October 2-6, 1995, Garching, Germany, Garching near Munich: European Southern Observatory,| c1996, edited by Martin Cullum, p. 95}, volume~54, page~95, 1996.

\bibitem{montera1997prediction}
Dennis~A Montera, Byron~M Welsh, Michael~C Roggemann, et~al.
\newblock Prediction of wave-front sensor slope measurements with artificial neural networks.
\newblock {\em Applied optics}, 36(3):675--681, 1997.

\bibitem{liu2020wavefront}
Xuewen Liu, Tim Morris, Chris Saunter, et~al.
\newblock Wavefront prediction using artificial neural networks for open-loop adaptive optics.
\newblock {\em Monthly Notices of the Royal Astronomical Society}, 2020.

\bibitem{swanson2021closed}
Robin Swanson, Masen Lamb, Carlos~M Correia, et~al.
\newblock Closed loop predictive control of adaptive optics systems with convolutional neural networks.
\newblock {\em Monthly Notices of the Royal Astronomical Society}, 503(2):2944--2954, 2021.

\bibitem{chen2021performance}
Justin~G Chen, Vinay Shah, and Lulu Liu.
\newblock Performance of a u-net-based neural network for predictive adaptive optics.
\newblock {\em Optics Letters}, 46(10):2513--2516, 2021.

\bibitem{haffert2021data}
Sebastiaan~Y Haffert, Jared~R Males, Laird~M Close, et~al.
\newblock Data-driven subspace predictive control of adaptive optics for high-contrast imaging.
\newblock {\em Journal of Astronomical Telescopes, Instruments, and Systems}, 7(2):029001--029001, 2021.

\bibitem{hafeez2022forecasting}
Rehan Hafeez, Finn Archinuk, S{\'e}bastien Fabbro, Hossen Teimoorinia, and Jean-Pierre V{\'e}ran.
\newblock Forecasting wavefront corrections in an adaptive optics system.
\newblock {\em Journal of Astronomical Telescopes, Instruments, and Systems}, 8(2):029003--029003, 2022.

\bibitem{wong2021predictive}
Alison~P Wong, Barnaby~RM Norris, Peter~G Tuthill, et~al.
\newblock Predictive control for adaptive optics using neural networks.
\newblock {\em Journal of Astronomical Telescopes, Instruments, and Systems}, 7(1):019001--019001, 2021.

\bibitem{hu2018build}
K~Hu, ZX~Xu, W~Yang, et~al.
\newblock Build the structure of wfsless ao system through deep reinforcement learning.
\newblock {\em IEEE Photonics Technology Letters}, 30(23):2033--2036, 2018.

\bibitem{ke2019self}
Hu~Ke, Bing Xu, Zhenxing Xu, et~al.
\newblock Self-learning control for wavefront sensorless adaptive optics system through deep reinforcement learning.
\newblock {\em Optik}, 178:785--793, 2019.

\bibitem{nousiainen2021adaptive}
Jalo Nousiainen, Chang Rajani, Markus Kasper, et~al.
\newblock Adaptive optics control using model-based reinforcement learning.
\newblock {\em Optics Express}, 29(10):15327--15344, 2021.

\bibitem{nousiainen2022towards}
Jalo Nousiainen, C~Rajani, M~Kasper, et~al.
\newblock Towards on-sky adaptive optics control using reinforcement learning.
\newblock {\em arXiv preprint arXiv:2205.07554}, 2022.

\bibitem{nousiainen2022advances}
Jalo Nousiainen, Byron Engler, Markus Kasper, Tapio Helin, C{\'e}dric~T Heritier, and Chang Rajani.
\newblock Advances in model-based reinforcement learning for adaptive optics control.
\newblock In {\em Adaptive Optics Systems VIII}, volume 12185, pages 882--891. SPIE, 2022.

\bibitem{nousiainen2024laboratory}
Jalo Nousiainen, Byron Engler, Markus Kasper, Chang Rajani, Tapio Helin, C{\'e}dric~T Heritier, Sascha~P Quanz, and Adrian~M Glauser.
\newblock Laboratory experiments of model-based reinforcement learning for adaptive optics control.
\newblock {\em Journal of Astronomical Telescopes, Instruments, and Systems}, 10(1):019001--019001, 2024.

\bibitem{landman2020self}
Rico Landman, Sebastiaan~Y Haffert, Vikram~M Radhakrishnan, et~al.
\newblock Self-optimizing adaptive optics control with reinforcement learning.
\newblock In {\em Adaptive Optics Systems VII}, volume 11448, page 1144849. International Society for Optics and Photonics, 2020.

\bibitem{landman2021self}
Rico Landman, Sebastiaan~Y Haffert, Vikram~M Radhakrishnan, et~al.
\newblock Self-optimizing adaptive optics control with reinforcement learning for high-contrast imaging.
\newblock {\em Journal of Astronomical Telescopes, Instruments, and Systems}, 7(3):039002, 2021.

\bibitem{pou2022model}
B~Pou, Jeffrey Smith, E~Quinones, et~al.
\newblock Model-free reinforcement learning with a non-linear reconstructor for closed-loop adaptive optics control with a pyramid wavefront sensor.
\newblock In {\em Adaptive Optics Systems VIII}, volume 12185, pages 945--958. SPIE, 2022.

\bibitem{pou2022adaptive}
B~Pou, Florian Ferreira, Eduardo Quinones, et~al.
\newblock Adaptive optics control with multi-agent model-free reinforcement learning.
\newblock {\em Optics express}, 30(2):2991--3015, 2022.

\bibitem{code_for_paper}
Bartomeu Pou, Jeffrey Smith, Eduardo Quinones, et~al.
\newblock Implementation of integrating supervised and reinforcement learning for predictive control with an unmodulated pyramid wavefront sensor for adaptive optics.
\newblock \url{https://github.com/Tomeu7/Integrating-SL-and-RL-for-AO}.

\bibitem{ferreira2018numerical}
Florian Ferreira, Eric Gendron, G{\'e}rard Rousset, et~al.
\newblock Numerical estimation of wavefront error breakdown in adaptive optics.
\newblock {\em Astronomy \& Astrophysics}, 616:A102, 2018.

\bibitem{ronneberger2015u}
Olaf Ronneberger, Philipp Fischer, and Thomas Brox.
\newblock U-net: Convolutional networks for biomedical image segmentation.
\newblock In {\em International Conference on Medical image computing and computer-assisted intervention}, pages 234--241. Springer, 2015.

\bibitem{he2016deep}
Kaiming He, Xiangyu Zhang, Shaoqing Ren, et~al.
\newblock Deep residual learning for image recognition.
\newblock In {\em Proceedings of the IEEE conference on computer vision and pattern recognition}, pages 770--778, 2016.

\bibitem{sutton2018reinforcement}
Richard~S Sutton and Andrew~G Barto.
\newblock {\em Reinforcement learning: An introduction}.
\newblock MIT press, 2018.

\bibitem{haarnoja2017reinforcement}
Tuomas Haarnoja, Haoran Tang, Pieter Abbeel, and Sergey Levine.
\newblock Reinforcement learning with deep energy-based policies.
\newblock In {\em International conference on machine learning}, pages 1352--1361. PMLR, 2017.

\bibitem{haarnoja2018soft}
Tuomas Haarnoja, Aurick Zhou, Kristian Hartikainen, et~al.
\newblock Soft actor-critic algorithms and applications.
\newblock {\em arXiv preprint arXiv:1812.05905}, 2018.

\bibitem{silver2018residual}
Tom Silver, Kelsey Allen, Josh Tenenbaum, et~al.
\newblock Residual policy learning.
\newblock {\em arXiv preprint arXiv:1812.06298}, 2018.

\bibitem{johannink2019residual}
Tobias Johannink, Shikhar Bahl, Ashvin Nair, et~al.
\newblock Residual reinforcement learning for robot control.
\newblock In {\em 2019 International Conference on Robotics and Automation (ICRA)}, pages 6023--6029. IEEE, 2019.

\bibitem{jovanovic2015subaru}
N~Jovanovic, Frantz Martinache, Olivier Guyon, Christophe Clergeon, Garima Singh, Tomoyuki Kudo, Vincent Garrel, Kevin Newman, D~Doughty, Julien Lozi, et~al.
\newblock The subaru coronagraphic extreme adaptive optics system: enabling high-contrast imaging on solar-system scales.
\newblock {\em Publications of the Astronomical Society of the Pacific}, 127(955):890, 2015.

\bibitem{roberts2011improved}
Lewis~C Roberts and L~William Bradford.
\newblock Improved models of upper-level wind for several astronomical observatories.
\newblock {\em Optics Express}, 19(2):820--837, 2011.

\bibitem{maas2013rectifier}
Andrew~L Maas, Awni~Y Hannun, Andrew~Y Ng, et~al.
\newblock Rectifier nonlinearities improve neural network acoustic models.
\newblock In {\em Proc. icml}, volume~30, page~3. Atlanta, GA, 2013.

\bibitem{ioffe2015batch}
Sergey Ioffe and Christian Szegedy.
\newblock Batch normalization: Accelerating deep network training by reducing internal covariate shift.
\newblock In {\em International conference on machine learning}, pages 448--456. PMLR, 2015.

\bibitem{kingma2014adam}
Diederik~P Kingma and Jimmy Ba.
\newblock Adam: A method for stochastic optimization.
\newblock {\em arXiv preprint arXiv:1412.6980}, 2014.

\bibitem{glorot2010understanding}
Xavier Glorot and Yoshua Bengio.
\newblock Understanding the difficulty of training deep feedforward neural networks.
\newblock In {\em Proceedings of the thirteenth international conference on artificial intelligence and statistics}, pages 249--256. JMLR Workshop and Conference Proceedings, 2010.

\bibitem{smith2023study}
Jeffrey~Peter Smith, Jesse Cranney, Charles Gretton, and Damien Gratadour.
\newblock A study of network-based wavefront estimation with noise.
\newblock In {\em Adaptive Optics for Extremely Large Telescopes 7th Edition}, 2023.

\bibitem{nvidia_tensorrt}
{NVIDIA Corporation}.
\newblock Tensorrt.
\newblock \url{https://developer.nvidia.com/tensorrt}, 2023.

\bibitem{plante2022high}
Julien Plante, Damien Gratadour, Lionel Matias, C{\'e}dric Viou, and Elena Agostini.
\newblock A high performance data acquisition on cots hardware for astronomical instrumentation.
\newblock In {\em Software and Cyberinfrastructure for Astronomy VII}, volume 12189, pages 323--332. SPIE, 2022.

\bibitem{bjorck2021low}
Johan Bj{\"o}rck, Xiangyu Chen, Christopher De~Sa, et~al.
\newblock Low-precision reinforcement learning: running soft actor-critic in half precision.
\newblock In {\em International Conference on Machine Learning}, pages 980--991. PMLR, 2021.

\bibitem{cobbe2021phasic}
Karl~W Cobbe, Jacob Hilton, Oleg Klimov, et~al.
\newblock Phasic policy gradient.
\newblock In {\em International Conference on Machine Learning}, pages 2020--2027. PMLR, 2021.

\bibitem{milli2018low}
Julien Milli, Markus Kasper, Pierre Bourget, Cyril Pannetier, David Mouillet, J-F Sauvage, Claudia Reyes, Thierry Fusco, Faustine Cantalloube, Konrad Tristam, et~al.
\newblock Low wind effect on vlt/sphere: impact, mitigation strategy, and results.
\newblock In {\em Adaptive Optics Systems VI}, volume 10703, pages 752--771. SPIE, 2018.

\bibitem{engler2019pyramid}
Byron Engler, Miska Louarn, Christophe Verinaud, Steve Weddell, and Richard Clare.
\newblock Pyramid wavefront sensing in the presence of thick spiders.
\newblock In {\em AO4ELT6}, 11 2019.

\bibitem{bertrou2020petalometry}
A~Bertrou-Cantou, E~Gendron, G~Rousset, F~Ferreira, A~Sevin, F~Vidal, Y~Cl{\'e}net, T~Buey, and S~Karkar.
\newblock Petalometry for the elt: dealing with the wavefront discontinuities induced by the telescope spider.
\newblock In {\em Adaptive Optics Systems VII}, volume 11448, pages 213--224. SPIE, 2020.

\bibitem{gendron1994astronomical}
Eric Gendron and Pierre L{\'e}na.
\newblock Astronomical adaptive optics. 1: Modal control optimization.
\newblock {\em Astronomy and Astrophysics (ISSN 0004-6361), vol. 291, no. 1, p. 337-347}, 291:337--347, 1994.

\bibitem{goulas2023saxo+}
Charles Goulas, Fabrice Vidal, Raphael Galicher, Johan Mazoyer, Florian Ferreira, Arnaud Sevin, Anthony Boccaletti, Eric Gendron, Cl{\'e}mentine B{\'e}chet, Michel Tallon, et~al.
\newblock Saxo+ upgrade: second stage ao system end-to-end numerical simulations.
\newblock {\em arXiv preprint arXiv:2310.15765}, 2023.

\end{thebibliography}

\end{document}